\documentclass[%
 aip,
 amsmath,amssymb,
 reprint,%
]{revtex4-2}
\usepackage[font=small,labelfont=bf]{caption}
\usepackage{overpic}
\usepackage[large]{subfigure }
\usepackage{amsmath,graphicx,color,epsfig,latexsym,bm,ulem,dutchcal,float,tikz,eucal, mathpazo,times, braket,comment}
\usepackage[colorlinks=true, linkcolor = blue, urlcolor  = blue, citecolor = blue, anchorcolor = red]{hyperref}   

\usepackage{lipsum}
\makeatletter
\newcommand{\shorteq}{%
	\settowidth{\@tempdima}{=}
	\resizebox{\@tempdima}{\height}{=}%
}
\makeatletter
\newcommand*\bigcdot{\mathpalette\bigcdot@{.5}}
\newcommand*\bigcdot@[2]{\mathbin{\vcenter{\hbox{\scalebox{#2}{$\m@th#1\bullet$}}}}}
\makeatother
\makeatother
\usepackage[right]{lineno}

\usepackage{caption}
\usepackage{subcaption}
\usepackage[svgnames]{xcolor}

\begin{document}
	
	
	\title{
    Weak Electron-Phonon Coupling Is Insufficient to Generate Significant CISS in Two-Terminal Transport}
\author{Vipul Upadhyay} \email{vipuupadhyay4@gmail.com}
	\affiliation{Department of Chemistry, Bar-Ilan University, Ramat-Gan 52900, Israel}
\affiliation{Institute of Nanotechnology and Advanced Materials, Bar-Ilan University, Ramat-Gan 52900, Israel}
\affiliation{Center for Quantum Entanglement Science and Technology, Bar-Ilan University, Ramat-Gan 52900, Israel}
	\author{Amikam Levy}\email{amikam.levy@biu.ac.il}
	\affiliation{Department of Chemistry, Bar-Ilan University, Ramat-Gan 52900, Israel}
\affiliation{Institute of Nanotechnology and Advanced Materials, Bar-Ilan University, Ramat-Gan 52900, Israel}
\affiliation{Center for Quantum Entanglement Science and Technology, Bar-Ilan University, Ramat-Gan 52900, Israel}

	\begin{abstract}{
A central open question in chiral-induced spin selectivity (CISS) is whether weak electron-phonon coupling in a helical molecular junction can generate a sizable spin polarization in two-terminal transport without invoking additional strong symmetry-breaking ingredients. We address this question by implementing a self-consistent nonequilibrium Green's function (NEGF) calculation for a helical tight-binding model with spin-orbit coupling and electron-phonon interactions. The electron-phonon self-energies are evaluated self-consistently, and the transport signal is extracted using the standard magnetization-reversal protocol with a spin-polarized analyzer lead.
We benchmark a fully self-consistent NEGF within the self-consistent Born approximation (SCBA) treatment for both global and local electron-phonon couplings against commonly used approximations, including diagonal self-energy schemes. We quantify how the resulting transport regime and spin polarization depend on phonon frequency, coupling strength, bias, temperature, and system size.
In contrast to large polarizations and anomalous size trends reported under approximate treatments, the fully self-consistent calculation yields negligible spin polarization, additionally the electron-phonon coupling mainly renormalizes the spectrum, and transport remains quasi-ballistic across the explored parameter range.
}
	\end{abstract}
	
	\maketitle
 
\section{Introduction}
The Chiral-Induced Spin Selectivity (CISS) effect, initially observed in transmission experiments \cite{naaman_exp_1Ray1999AsymmetricScattering,Gohler2011,exp_Xie2011,exp_Mondal2016,exp_misrapmid23980184}, refers to the spin-based filtering or selectivity of electrons as they pass through chiral molecules, occurring even in the absence of external magnetic fields.
Remarkably, reported spin polarizations can reach tens of percent~\cite{naaman_exp_1Ray1999AsymmetricScattering,Gohler2011}, motivating potential applications, particularly in spintronics~\cite{Review_doi:10.1021/acs.chemrev.3c00661,exp_Mondal2016,Naaman2020CISS,Review_Gupta_D4SC05736H}.
From the moment the CISS effect was experimentally observed, theoretical efforts have been made to understand its microscopic origins \cite{guttirrez_PhysRevB.85.081404,Guo_PhysRevLett.108.218102,thery_ciss_https://doi.org/10.1002/adma.202106629,early_thoery_ciss_10.1063/1.3167404}. Spin-orbit coupling (SOC) emerged as a promising candidate to explain the phenomenon, due to its ability to mix spins of different orientations. However, because the CISS effect is typically observed in organic molecules, such as DNA, the SOC in these systems is generally weak and insufficient to explain the substantial spin polarization observed experimentally. As such, no widely accepted microscopic theory for CISS exists \cite{Review_doi:10.1021/acs.chemrev.3c00661,thery_ciss_https://doi.org/10.1002/adma.202106629}. 

\par Any theoretical model considered should address the single-channel no-go theorem~\cite{Guo_PhysRevLett.108.218102,thery_ciss_https://doi.org/10.1002/adma.202106629}, which states that in strictly one-dimensional single-channel systems the SOC term can be gauged out of the Hamiltonian in the absence of spin-dependent potentials; therefore, SOC alone cannot produce robust spin-polarized transmission in such a minimal model.
Furthermore, from a symmetry perspective, it is important to note that chirality breaks spatial inversion but does not, by itself, violate time-reversal symmetry (TRS) \cite{BARRON1986423}. Consequently, chirality alone cannot guarantee spin-selective transport in two-terminal setups, where Onsager reciprocity~\cite{onsager1931reciprocal} and Bardarson theorem \cite{Bardarson_2008}, enforce equal conductance for opposite spin channels in the absence of explicit TRS breaking~\cite{casimir1945onsager}. In many CISS experiments, however, spin-resolved signals are measured using magnetic or spin-projective contacts \cite{Fransson_What_it_CISS_https://doi.org/10.1002/ijch.202200046}, which already break TRS at the level of the leads or detectors. Within such setups, chirality can enable spin-selective transmission without violating global TRS, a viewpoint that underlies several recent theoretical approaches\cite{scipost_10.21468/SciPostPhysCore.6.2.044,Aharony10.1063/5.0265363,Aharony_Ciss_PhysRevB.102.035445,fransson2025breakingtimereversalsymmetryonsager,dalum_doi:10.1021/acs.nanolett.9b01707}. 
One of the earlier theoretical studies based on these considerations was conducted by Guo et al. \cite{Guo_PhysRevLett.108.218102}, who observed relatively large spin polarization values on working with virtual B\"{u}ttiker baths, \cite{probe_dvira_10.1063/1.4926395}. However, the spin-orbit coupling (SOC) strength used in their model was 1-2 orders of magnitude higher than experimentally observed values. More recent approaches have explored various other methods to amplify the SOC strength, such as incorporating a feedback mechanism into the Schr\"{o}dinger equation \cite{feedback2_PhysRevB.104.024430}, considering CISS as an interface effect \cite{feedback_doi:10.1021/jacs.1c05637,sarkar2025spinterface,Dubi_Temp_10.1063/5.0147886}, or attributing the amplification in spin { polarization} to nuclear motion \cite{wenjie_nuclear_PhysRevB.106.184302,Nuclear2PhysRevB.105.195117}. 

\par A growing body of work suggests that sizeable CISS signals are difficult to account for within purely coherent single-particle descriptions, and may require additional ingredients beyond spin-orbit coupling alone, such as many-body interactions, environmental dephasing, or interfacial effects~\cite{Review_doi:10.1021/acs.chemrev.3c00661,Fransson_electron_correlation_doi:10.1021/acs.jpclett.9b02929,Fransson_What_it_CISS_https://doi.org/10.1002/ijch.202200046,many_body_Chiesa2024,Fransson_latest_10.1063/5.0289548,Takehito_10.1063/5.0207915}
A central open question in CISS is whether weak electron-phonon (e-ph) coupling can generate a sizable spin polarization in two-terminal transport without additional strong symmetry-breaking ingredients~\cite{Fransson_PhysRevB.102.235416,wenjie_CJCP2509149,Thoss}. This vibrational mechanism was predicted to amplify spin polarization by modifying the density of states for oppositely magnetized leads~\cite{Fransson_PhysRevB.102.235416}. However, these predictions were obtained using a non-self-consistent treatment of the interaction self-energies, which can violate conservation laws (e.g., current conservation) and introduce unphysical spectral features. Recent studies have revisited this mechanism using alternative approaches, including hierarchical equations of motion~\cite{Thoss} and semiclassical treatments~\cite{wenjie_CJCP2509149}, and reported that large polarization may appear only for individual trajectories~\cite{Thoss} or during transient dynamics~\cite{wenjie_CJCP2509149}, while ensemble- or time-averaged steady-state polarization remains negligible.

Here we provide a controlled weak-coupling benchmark by solving the nonequilibrium Green's-function equations with e-ph self-energies evaluated fully self-consistently within the self-consistent Born approximation (SCBA)~\cite{e-ph_PhysRevB.72.201101}. We consider both global and local e-ph couplings and explicitly compare the converged self-consistent solution to commonly used approximate schemes, including diagonal self-energy treatments. This enables a direct assessment of how approximation choices impact the transport regime and the inferred spin polarization.
{ In addition, motivated by the recent suggestion that vibrational relaxation due to an external thermal bath may be important for CISS, we also examine a phenomenological finite-phonon-linewidth extension within the same self-consistent framework and assess whether it qualitatively changes the resulting transport spin polarization.}

The manuscript is organised as follows. We start by describing the model in the section \ref{model}. We then discuss the relevant self-consistent NEGF equations used for analysing the model in the section \ref{sec:NEGF_Equations}. This is followed by detailed discussion on obtained results, first the Energy-dependent results in the section \ref{sec:results}, followed by the energy integrated results in the section \ref{Sec:Results:B}. Finally, we conclude in the section \ref{sec:conclusion}. A detailed derivation of all the equation used for the analysis can be found in the appendix \ref{deriv_eqns}.

	\section{Model for vibrationally amplified CISS effect}  \label{model}
\begin{figure}[t]
		\centering 
		{\includegraphics[width=1\linewidth]{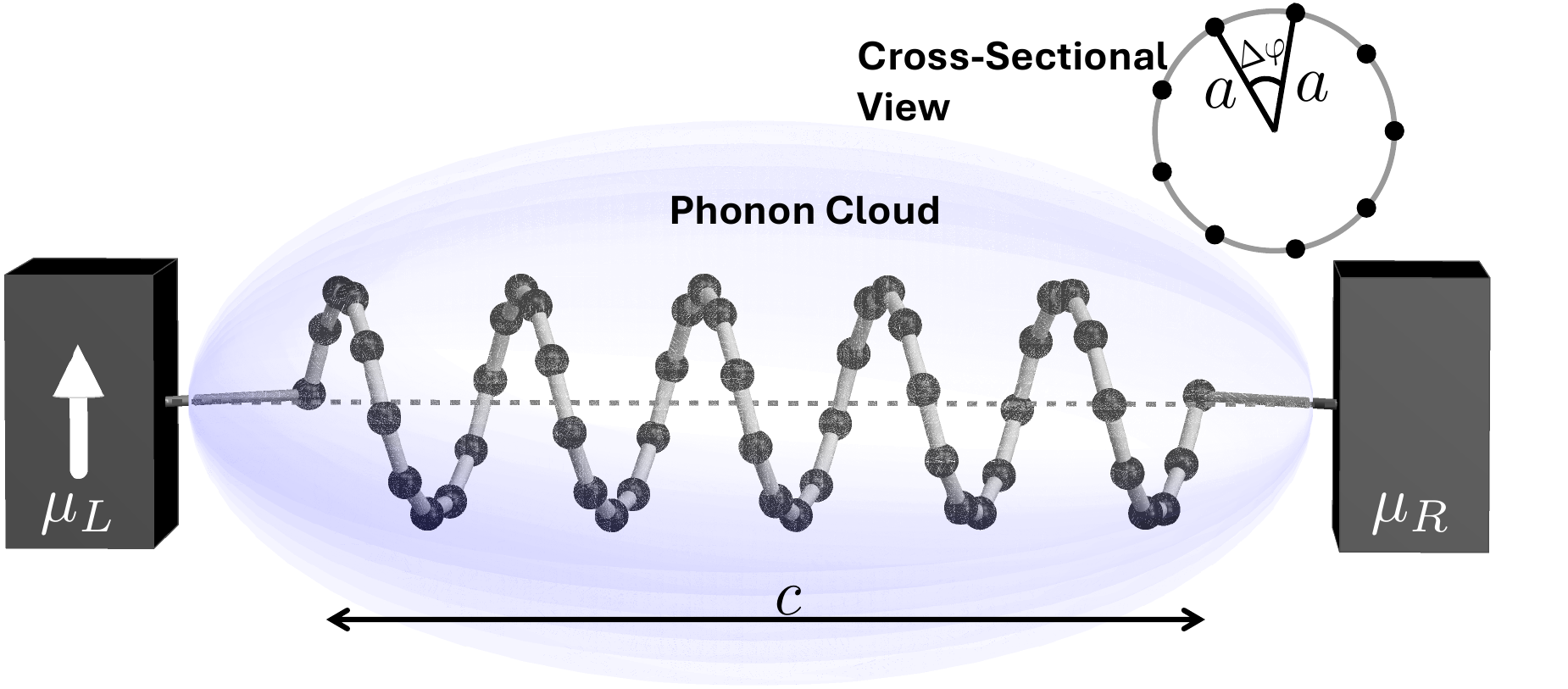}}
		\caption{Schematic of the theoretical setup. A helical molecule is kept between two leads with chemical potentials, $\mu_L = V/2$ and $\mu_R = -V/2$. The left lead is magnetised one way and the particle current is measured. The magnetization is then reversed and the current is measured again. The difference between the two currents quantifies the CISS-induced spin polarization. Specific parameters for this figure are $n_l=10, m_l=5$, radius of the helix $a=1$, and length of the molecule $c=8$.}

		\label{Model_fig}
	\end{figure}

We begin by outlining Fransson's model for CISS  (see Fig.~\ref{Model_fig}). This model is based on a tight-binding Hamiltonian including chiral features, which are introduced due to the helical nature of the structure. This Hamiltonian is then augmented with spin-dependent electron-vibration interactions, which are indicated to enhance the spin polarization in the system. The Hamiltonian is explicitly written as \cite{Fransson_PhysRevB.102.235416},
\begin{align}\label{H_mol}
&\mathcal{H}_{\text{mol}} =
\sum_{m=1}^{M}  \psi_m^\dagger \psi_m\left[ \varepsilon_0+ \sum_{\nu}\varepsilon_\nu (a_{\nu} + a_{\nu}^\dagger) \right] +
\sum_\nu\omega_{\nu} a_{\nu}^\dagger a_{\nu} 
\nonumber\\&-\sum_{m=1}^{M-1} \left( \psi_m^\dagger \psi_{m+1} + \text{H.c.} \right)
\left[ t_0 + \sum_\nu t_{\nu} (a_{\nu} + a_{\nu}^\dagger) \right] \nonumber\\&
+\sum_{m=1}^{M-2} \left( i \psi_m^\dagger \boldsymbol{v_m^{(+)}} \cdot \boldsymbol{\sigma} \psi_{m+2} + \text{H.c.} \right)
\left[ \lambda_0 + \sum_\nu \lambda_{\nu} (a_{\nu} + a_{\nu}^\dagger) \right].
\end{align}
Here, $\varepsilon_0$ denotes the on-site energy of the electronic state at site $m$, and $\psi_m^\dagger$ ($\psi_m$) are the creation (annihilation) operators for an electron at site $m$, with each $\psi_m$ being a two-component spinor representing up and down spin. The vibrational modes are characterized by frequencies $\omega_\nu$, with $a_\nu^\dagger$ and $a_\nu$ denoting the phonon creation and annihilation operators, respectively. The parameter $t_0$ represents the nearest-neighbor hopping amplitude, while $t_\nu$ describes the e-ph-assisted hopping for mode $\nu$. The spin-orbit coupling is captured by $\lambda_0$, with $\lambda_\nu$ representing the e-ph-assisted spin-orbit term. { To generalise the above model, we also modify it by adding a local on site e-ph interaction similar to the Holstein interaction \cite{Galperin_2007,HOLSTEIN1959325}, quantified by the parameter $\varepsilon_\nu$.}
\par The vector $\boldsymbol{v_m^{(+)}}$ defines the chirality of the spin-orbit interaction between next-nearest neighbors, and $\boldsymbol{\sigma}$ is the vector of Pauli matrices representing electron spin, with H.c. denoting the Hermitian conjugate.  We work in the natural units so $\it{e}=1,\hbar=1,k_B=1$. The chirality vector is defined as
\begin{equation}
\boldsymbol{v_m^{(+)}} = \boldsymbol{\hat{d}_{m,1} \times \hat{d}_{m+1,1}},
\end{equation}
where $\boldsymbol{\hat{d}_{m,s}}$ is the unit vector along the bond connecting sites $m$ and $m+s$, given by
\begin{equation}
\boldsymbol{\hat{d}_{m,s}} = \frac{\boldsymbol{r_{m+s}} - \boldsymbol{r_m}}{|\boldsymbol{r_{m+s}} - \boldsymbol{r_m}|}.
\end{equation}
The positions of the sites in the helical structure are specified as,

\begin{equation}
\boldsymbol{r_m} = \left( a \cos \varphi_m, a \sin \varphi_m, \frac{(m-1)c}{M-1} \right),
\end{equation}
where $a$ is the radius of the helix, $c$ is the length of the helix, $M$ is the total number of sites, and $\varphi_m = \frac{2\pi (m-1) m_l}{M-1}$ is the azimuthal angle of site $m$, with $\Delta \varphi=\frac{2\pi m_l}{M-1}$, being the angular separation between two neighboring site. The helix consists of $M = m_l \times n_l$ sites, where $m_l$ is the number of loops and $n_l$ is the number of sites per loop (see Fig.~\ref{Model_fig}). 
The analysis is carried out using the Nonequilibrium Green's Function (NEGF) under SCBA approach discussed below. These equations are derived using the Quantum Langevin Equation (QLE) approach \cite{Dhar_PhysRevB.73.085119,Datta_2005}. An explicit derivation of the used equation is given in the appendix \ref{deriv_eqns}. 

\section{Nonequilibrium Green's Function Equations} \label{sec:NEGF_Equations}
The QLE method is based on separating the different components of the dynamical equation of motion into dissipative and noise parts \cite{Dhar_PhysRevB.73.085119}. Once this is achieved, it becomes possible to calculate currents and density of states \cite{Dhar_PhysRevB.73.085119,Datta_2005}. For the given model, we have two real leads in the model through which the current is calculated using the usual Landauer-B\"{u}ttiker formalism \cite{landauer1957spatial,buttiker1986four,Datta_2005}. However, once we include the interactions with the environment using SCBA, the equations modify, and we need to solve them self-consistently. This involves effectively making a new virtual lead for e-ph interactions. The properties of this lead depend on the Green's Functions of the system, so we need to do self-consistent evaluations. We start by defining the retarded Green's Function $G^R$, which is the central quantity in any steady state NEGF analysis. 
For our case, the retarded Green's Function is,
\begin{align}\label{slef_consitent_1}
G^R(E)=\Big[E-H+\frac{i}{2}\big(\Gamma_L+\Gamma_R\big)-\Sigma_{\mathrm{ph}}^{R}(E)\Big]^{-1},
\end{align}
where we assume a wide-band (flat-spectrum) approximation for the real leads~\cite{Fransson_PhysRevB.102.235416}.
We define the phonon-induced broadening matrix by
\begin{align}
\Gamma_{\mathrm{ph}}(E)\equiv i\left[\Sigma_{\mathrm{ph}}^{R}(E)-\Sigma_{\mathrm{ph}}^{A}(E)\right],
\end{align}
which implies the standard Hilbert-transform relation~\cite{Bijay_negf_PhysRevLett.130.187101}
\begin{align}\label{selconsis_2}
\Sigma_{\mathrm{ph}}^{R}(E)
&=\frac{1}{2\pi}\mathcal{P}\!\int_{-\infty}^{\infty}\! dE'\,
\frac{\Gamma_{\mathrm{ph}}(E')}{E-E'}
-\frac{i}{2}\Gamma_{\mathrm{ph}}(E).
\end{align}

We assume the phonons are in thermal equilibrium at temperature $T$, i.e., their occupations are fixed to the
Bose-Einstein distribution $n_B(\omega,T)=(e^{\omega/T}-1)^{-1}$ (no nonequilibrium phonon heating).
For compactness we define $n_\gamma\equiv n_B(\omega_\gamma,T)$ and use the shorthand
$G^{</>}_{\pm\gamma}(E)\equiv G^{</>}(E\pm\omega_\gamma)$.
The phonon broadening is then (see Appendix~\eqref{broadening_appendix})\cite{e-ph_PhysRevB.72.201101}
\begin{align}\label{Gamma_ph_explicit_pm}
\Gamma_{\mathrm{ph}}(E)
&= i \sum_\gamma M^\gamma\Big[
(1+n_\gamma)\,G^{>}_{-\gamma}(E)-n_\gamma\,G^{<}_{-\gamma}(E)
\Big]M^\gamma \nonumber\\
&\quad + i \sum_\gamma M^\gamma\Big[
n_\gamma\,G^{>}_{+\gamma}(E)-(1+n_\gamma)\,G^{<}_{+\gamma}(E)
\Big]M^\gamma ,
\end{align}
where $M^\gamma$ is the electron-phonon coupling matrix determined by the parameters $\varepsilon_\gamma,t_\gamma,\lambda_\gamma$.
Once $G^R(E)$ is known, the density of available states follows as,
\begin{align}\label{density_of_states}
\rho(E)
&=\frac{i}{2\pi}\mathrm{Tr}\!\left[G^R(E)-G^A(E)\right]
=\frac{i}{2\pi}\mathrm{Tr}\!\left[G^>(E)-G^<(E)\right],
\end{align}
with $G^A(E)=G^R(E)^\dagger$.
To compute particle currents we also require the occupied states,
\begin{align}\label{slef_consitent_3}
G^<(E)
&= i\,G^R(E)\Big[f_L(E)\Gamma_L+f_R(E)\Gamma_R\Big]G^A(E) \nonumber\\
&\quad + G^R(E)\Sigma_{\mathrm{ph}}^<(E)G^A(E),
\end{align}
where  $f_\alpha(E)=[e^{(E-\mu_\alpha)/T}+1]^{-1}$ is the Fermi-Dirac distribution of lead $\alpha$ and the phonon in-scattering self-energy is (see Appendix~\eqref{inscattering_appenidx})\cite{e-ph_PhysRevB.72.201101}
\begin{align}\label{self_consi_4}
\Sigma_{\mathrm{ph}}^<(E)
&=& \sum_\gamma M^\gamma\big[
n_\gamma\,G^{<}_{-\gamma}(E)
+\big(1+n_\gamma\big)\,G^{<}_{+\gamma}(E)
\big]M^\gamma .
\end{align}
Finally, the net particle current (incoming and outgoing) flowing out of lead $\alpha$ 
can be written in the interacting NEGF form~\cite{meir1992landauer,jauho1994time}, which we evaluate using the self-consistent Green’s functions:

\begin{align}\label{current_definition}
I_\alpha
&=\frac{i}{2\pi}\int_{-\infty}^{\infty}\! dE\;
\mathrm{Tr}\!\left\{\Gamma_\alpha
\Big[f_\alpha\,G^>(E)+\big(1-f_\alpha\big)\,G^<(E)\Big]\right\}.
\end{align}

The CISS measurements are typically performed by comparing two transport experiments in which the spin polarization of one of the leads is reversed \cite{Fransson_What_it_CISS_https://doi.org/10.1002/ijch.202200046}. In our theoretical description, this procedure is implemented by modifying the spin-dependent broadening matrix of the left lead. Specifically, the non-zero elements are taken as,
\begin{equation}
\Gamma^L_{\uparrow,\downarrow}=\Gamma_0\,(1\pm p),
\end{equation}
where the polarization parameter takes opposite values, \(p=\pm 0.5\), in the two experiments. The spin polarization is then extracted by comparing the resulting steady-state currents for the two lead polarizations, using the formula,
\begin{align}
\text{P}_\text{T}=100\times \frac{I^L_\uparrow-I^L_\downarrow}{I^L_\uparrow+I^L_\downarrow}
\end{align}
 For simplicity, the present analysis is restricted to a single phonon mode \cite{Fransson_PhysRevB.102.235416,Thoss,wenjie_CJCP2509149}, so $\varepsilon_\nu=\varepsilon_1,t_\nu=t_1,\lambda_\nu=\lambda_1,$ and $ \omega_\nu=\omega_0$.
We now solve the above self-consistent equations to determine different physical quantities, and ultimately to calculate the spin polarization.

{ \par It is important to clarify what is meant here by ``weak'' electron-phonon coupling. In the present NEGF--SCBA framework, the interaction enters through the electronic self-energies $\Sigma_{\rm ph}^{R/<}(E)$ [Eqs.~\eqref{selconsis_2} and \eqref{self_consi_4}], which are evaluated at Born level, i.e., to second order in the electron-phonon coupling matrix $M^\gamma$, but self-consistently with the dressed electronic Green's functions. As discussed in earlier studies, this approximation is quantitatively reliable only when the electron-phonon self-energy remains a perturbative correction to the dominant electronic scales of the junction \cite{Weak_SCBAPhysRevB.79.085120,Galperin_2007}.

A convenient dimensionless measure of this regime is the ratio between the characteristic phonon-induced dressing scale and the electronic escape rate into the leads \cite{Galperin_2007}. For a single mode of frequency $\omega_0$ and a representative lead-induced broadening $\Gamma$ (here taken as $\Gamma\simeq\Gamma_0$), this gives
\begin{equation}
g_{\rm ep}\sim\frac{\|M^\gamma\|^2}{\omega_0\,\Gamma}.
\end{equation}
In our helical model, the coupling matrix $M^\gamma$ is determined by the parameters $(\epsilon_1,t_1,\lambda_1)$, where $\epsilon_1$ and $t_1$ describe spin-independent electron-phonon couplings (onsite and hopping-modulation terms, respectively), while $\lambda_1$ describes the spin-dependent SOC-assisted coupling. It is therefore useful to define
\begin{equation}
g_t=\frac{t_1^2}{\omega_0\Gamma_0},
\qquad
g_\lambda=\frac{\lambda_1^2}{\omega_0\Gamma_0},
\end{equation}
and, when the onsite term is present,
\begin{equation}
g_\epsilon=\frac{\epsilon_1^2}{\omega_0\Gamma_0}.
\end{equation}

At finite temperature, the phonon-induced electronic broadening is further weighted by the Bose factors $n_B(\omega_0,T)$ and $1+n_B(\omega_0,T)$ appearing in Eqs.~\eqref{selconsis_2} and \eqref{self_consi_4}. Consequently, for low phonon frequencies the effective scattering can be substantially enhanced because $n_B(\omega_0,T)$ becomes large when $\omega_0\ll T$. In the following, by ``weak e--ph coupling'' we therefore mean the regime in which the phonon-induced electronic self-energy remains a perturbative correction to the lead-induced broadening, i.e.\ $\|\Gamma_{\rm ph}\|\ll \|\Gamma_L+\Gamma_R\|$, so that SCBA is expected to remain quantitatively reliable.}

\subsection{Details about numerical analysis} \label{numerical_details}
As evident from the previous section, the correct analysis of the model requires solving Eqs. \eqref{slef_consitent_1}–\eqref{self_consi_4} self-consistently. These coupled equations form a nonlinear, energy-dependent problem and are, in general, numerically challenging to solve \cite{numerical_inealstic_PhysRevB.75.205413}. 
To this end, we discretize the energy domain on a uniform grid with a spacing equal to the phonon frequency. This choice ensures that Green’s functions connected via phonon absorption and emission processes are treated in a mutually consistent manner. The Green’s functions and self-energies are obtained self-consistently through an iterative procedure until convergence is reached \cite{numerical_inealstic_PhysRevB.75.205413}. Since the total number of grid points is determined by the phonon frequency, we repeatedly shift the energy grid to effectively enhance the energy resolution and capture finer spectral features.

\par To improve convergence and suppress numerical instabilities, Pulay mixing algorithm \cite{PULAY1980393} is employed, wherein a linear combination of Green’s functions from several previous iterations is used to construct the next iterate. Convergence is monitored via the relative Frobenius norm of the residual between successive Green’s functions, $\frac{|| G^R_{New}-G^R_{old}||}{||G^R_{New}||}$, with the tolerance set to $5\times10^{-5}$. In Sections \ref{sec:results} and \ref{Sec:Results:B}, we verify the physical validity of our analysis through several consistency checks, such as particle current conservation and the vanishing of current at zero bias.
\par
Furthermore, the spin-polarization plots are obtained while explicitly accounting for residual numerical noise arising from the finite self-consistency tolerance. This noise is considered to be proportional to the sum of left, right, and phonon lead currents, which in the ideal case should give us zero. This tolerance leads to the appearance of sharp cutoffs in the polarization plots.

\section{Energy-dependent Results}\label{sec:results}

\begin{figure*}[t]
\centering

\begin{overpic}[width=0.32\linewidth]{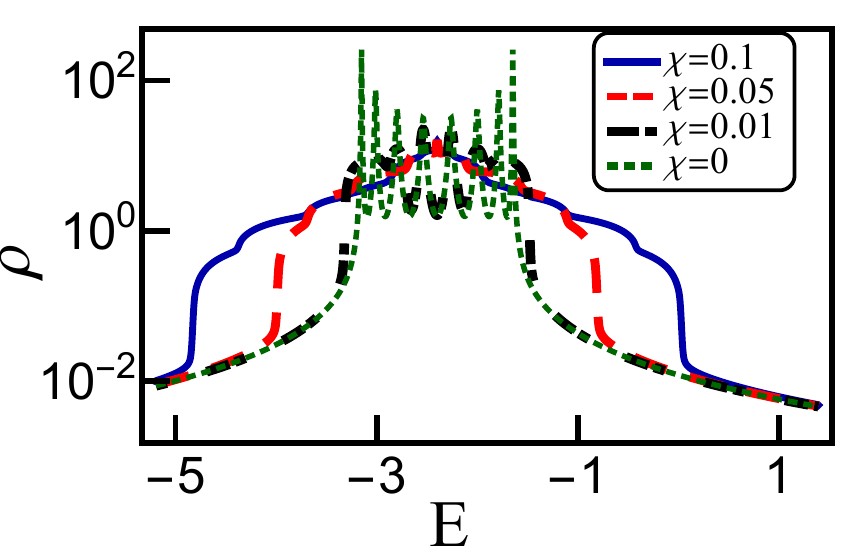}
\put(20,50){ \large\bfseries{(a)}}
\end{overpic}
\begin{overpic}[width=0.32\linewidth]{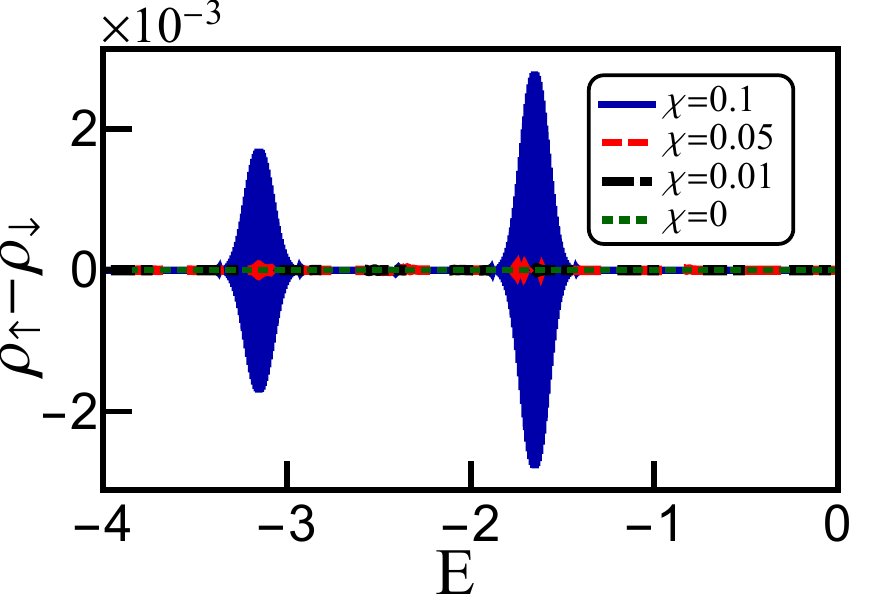}
\put(15,55){ \large\bfseries{(b)}}
\end{overpic}
\begin{overpic}[width=0.32\linewidth]{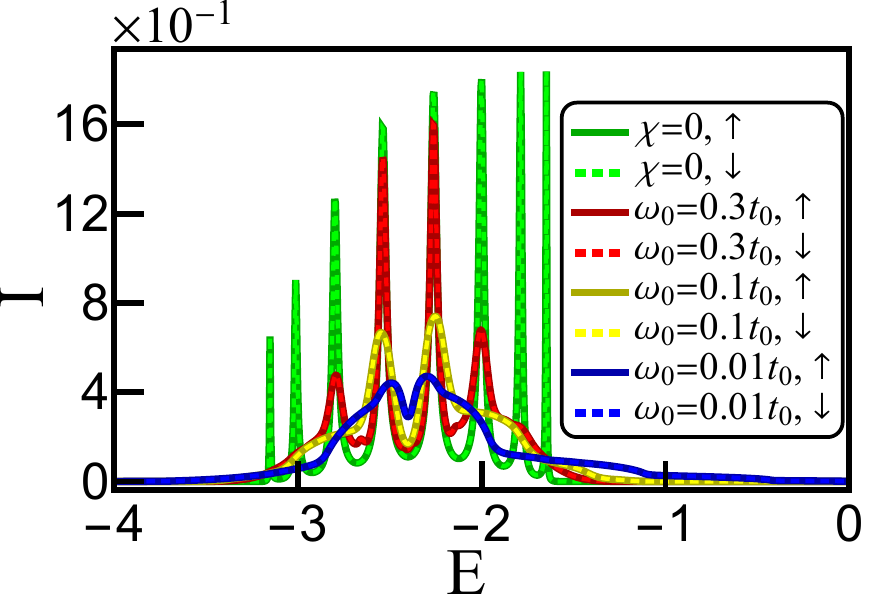}
\put(15,55){ \large\bfseries{(c)}}
\end{overpic}
\begin{overpic}[width=0.32\linewidth]{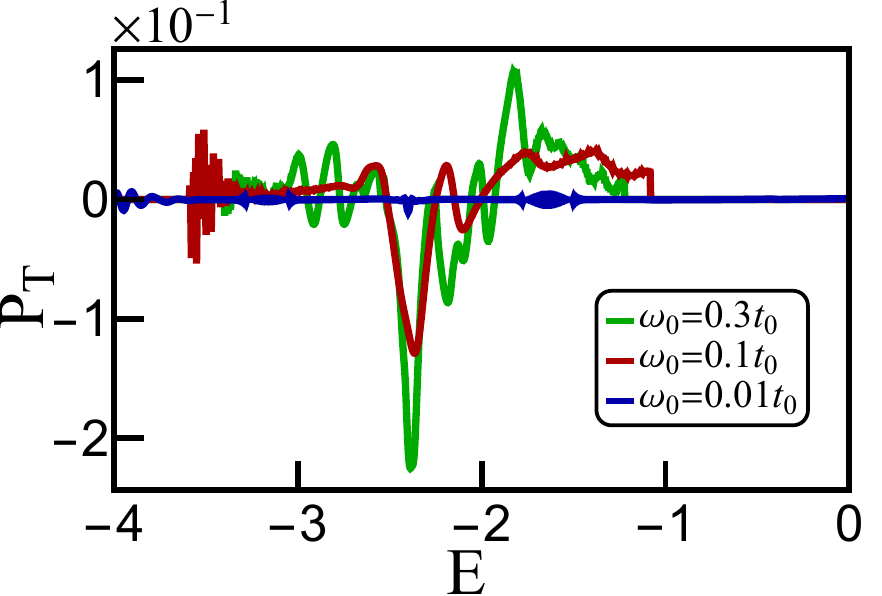}
\put(15,55){ \large\bfseries{(d)}}
\end{overpic}
\begin{overpic}[width=0.32\linewidth]{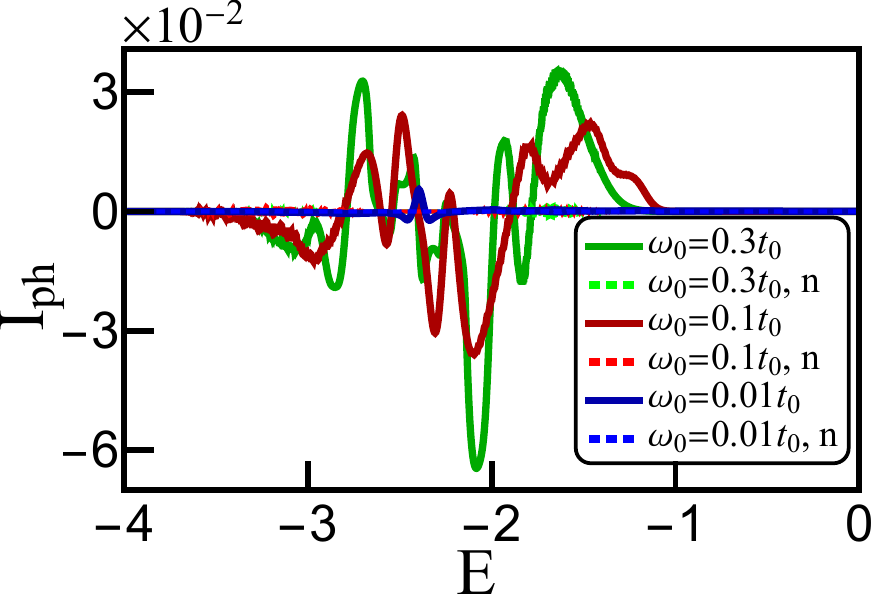}
\put(15,55){ \large \bfseries{(e)}}
\end{overpic}
\begin{overpic}[width=0.32\linewidth]{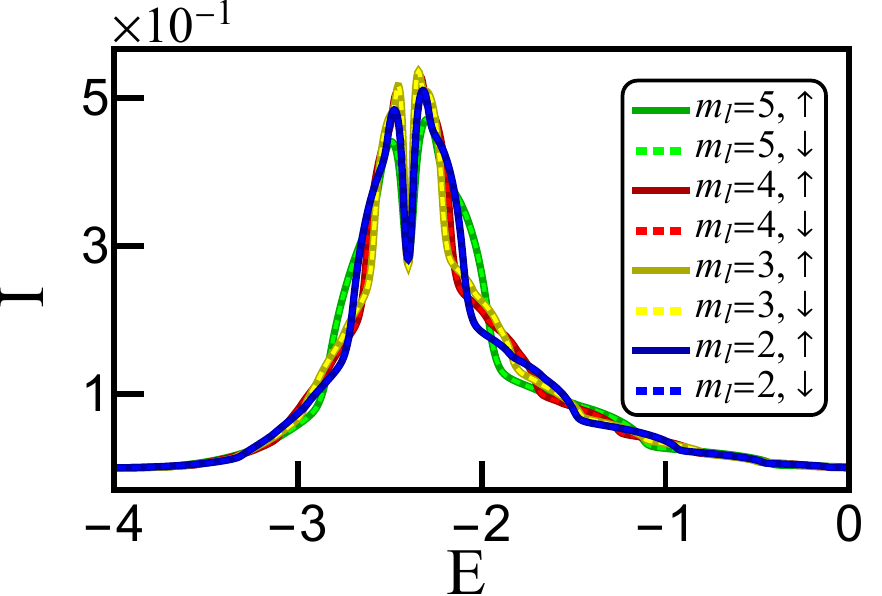}
\put(15,55){ \large\bfseries{(f)}}
\end{overpic}

\caption{\textbf{Energy-dependent Figures with Global e-ph interaction:}
\textbf{(a)} Variation of density of states for different e-ph interaction strength $\chi$, such that $t_1=\chi t_0, \lambda_1=\chi \lambda_0$.
\textbf{(b)} Difference in density of electronic states for opposite lead magnetization, $\rho_\uparrow-\rho_\downarrow$, for different e-ph interaction strengths. 
\textbf{(c)} Variation of particle current coming out of the left bath, for opposite lead magnetization and different phonon frequencies. 
\textbf{(d)} Energy-dependent spin polarization for different phonon frequencies. 
\textbf{(e)} Energy-dependent phonon-bath current and numerical current noise denoted by the symbol `n' for different phonon frequencies and left lead polarisation, $p=0.5$. The numerical noise is defined in subsection \ref{numerical_details}.
\textbf{(f)} Energy-dependent current for different system sizes. The default system size considered above is $n_l=4, m_l=2$.  
The e-ph coupling is similar to the study \cite{Fransson_PhysRevB.102.235416}, with $\varepsilon_1=0$, and
the default parameter values in units of $0.1$ eV are $t_0=0.4,\varepsilon_0=-6t_0, t_1=0.1t_0, \lambda_0=t_0/40,$ $\lambda_1=0.1\lambda_0$, $\Gamma_0=t_0/4, \omega_0=0.01t_0, T=0.25$ $(300\,\text{K})$. Also, radius and length of helix are equal, $a=c=1$. The chemical potentials are $\mu_L=V/2, \mu_R=-V/2$, with $V=15t_0$.
}
\label{ideal_fig}
\end{figure*}

\subsection{Global Interaction}
We first discuss the results obtained when analysing Fransson's model \cite{Fransson_PhysRevB.102.235416}, but solving it self-consistently. This means that the onsite e-ph interaction is turned off, $\epsilon_1=0$, and we only have hopping assistance due to the interactions. We call this the global (G) e-ph interaction model. 
\par We start the analysis by studying the energy dependence of the density of states (DOS), shown in Fig. \ref{ideal_fig}(a), calculated using Eq. \eqref{density_of_states}. In the absence of interactions, the DOS exhibits several well-defined peaks corresponding to the eigenvalues of the non-interacting Hamiltonian. The finite width of these peaks arises from the coupling to the real leads, which introduces lifetime broadening and results in a continuous spectrum. 
When e-ph interactions are included, the DOS undergoes a marked qualitative change. Quantifying the strength of the e-ph interaction by $\chi$, such that $t_1=\chi t_0, \lambda_1=\chi \lambda_0$, we see in Fig. \ref{ideal_fig}(a) that as soon as the interactions are turned on, the peaks become broader while simultaneously their maximum value decreases. This trend keeps up with increasing $\chi$, and for $\chi\sim 0.1$, the sharp resonant peaks present in the non-interacting case are strongly suppressed, and the DOS evolves into a much smoother profile. In particular, for larger e-ph interaction, a broad plateau develops in the central energy region, with the spectral density gradually decreasing on either side. Compared to the non-interacting case, the overall DOS is significantly broadened, indicating that electron-phonon coupling leads to a substantial redistribution of spectral weight over energy.
This overall trend is partially similar to the behavior reported in the study \cite{Fransson_PhysRevB.102.235416}, where electron-phonon interactions also lead to significant modification of the spectral features. However, there are important qualitative differences between the two treatments. In the earlier study, the individual resonant peaks remain clearly visible even in the presence of stronger interactions (see Fig.\ref{Fransson_method_fig}(a) in Appendix \ref{Fransson_results}), whereas in our fully self-consistent calculation, these peaks are largely washed out. This distinction suggests that the degree of self-consistency and the treatment of interactions play a crucial role in determining the resulting spectral properties.

{  \par As discussed in Sec.~\ref{sec:NEGF_Equations}, the SCBA equations used here are quantitatively reliable only when the relevant dimensionless coupling strengths remain sufficiently small. In the present global-interaction model, this distinction is particularly important because the spin-dependent and spin-independent electron--phonon channels play different roles. The spin-dependent coupling, governed by $\lambda_1$, remains weak throughout the parameter range considered, since $g_\lambda=\lambda_1^2/(\omega_0\Gamma_0)\ll 1$. By contrast, the spin-independent coupling associated with $t_1$ need not remain weak when the phonon frequency is very small.

In particular, for the lowest phonon frequency considered here, $\omega_0=0.01t_0$, the parameter
\begin{equation}
g_t=\frac{t_1^2}{\omega_0\Gamma_0}
\end{equation}
is no longer much smaller than unity. In this low-frequency regime, the phonon-induced electronic broadening is additionally enhanced by the Bose factors entering the SCBA self-energies, and therefore the large broadening and smoothing of the energy-dependent spectra seen in Fig.~\ref{ideal_fig}(a) should not be interpreted as a quantitatively controlled weak-coupling prediction. Rather, in this corner of parameter space, SCBA should be viewed as providing qualitative trends only.

For larger phonon frequencies, however, the ratio $t_1^2/(\omega_0\Gamma_0)$ decreases, the phonon-induced broadening becomes smaller relative to the lead-induced broadening, and the weak-coupling conditions underlying SCBA are better satisfied. The inclusion of the very small phonon frequencies in Fig.~\ref{ideal_fig} is retained mainly to enable direct comparison with Ref.~\cite{Fransson_PhysRevB.102.235416}, where a similar low-frequency parameter regime was considered.}

\par To explore the implications of these changes for spin-dependent effects, we next plot the difference in the DOS for opposite lead magnetizations in Fig. \ref{ideal_fig}(b). The resulting difference remains small across the entire energy range and does not show pronounced features that would indicate favorable conditions for the emergence of significant spin polarization. This is an interesting result, as it contrasts with the conclusions drawn in earlier studies \cite{Fransson_PhysRevB.102.235416,fransson2025chiralinducedspinpolarized}, where larger spin-dependent DOS differences were reported.
\par We then examine the transport properties by analyzing the energy-resolved particle current flowing out of the left lead, evaluated using the integrand of Eq. \eqref{current_definition}. This energy-dependent particle current for different phonon frequencies is shown in Fig.~\ref{ideal_fig}(c). We observe that with increasing phonon frequency, the peak magnitude of the particle current increases, accompanied by progressively sharper and more well-defined spectral peaks.
{This is because at lower phonon frequencies, the
self-consistent electron-phonon self-energy leads to stronger effective spectral broadening, so the discrete resonances are smeared
and the current profile becomes smoother. At higher phonon frequencies, this broadening is weaker, and the spectral features of
the underlying Hamiltonian remain more clearly resolved.}
However, the energy-dependent lead currents corresponding to opposite lead magnetizations are visually indistinguishable over the entire energy window considered for the different phonon frequencies. This absence suggests that reversing the lead magnetization has a negligible effect on the transport characteristics, which serves as an important benchmark for a CISS setup.

\begin{figure*}[t]
\centering

\begin{overpic}[width=0.32\linewidth]{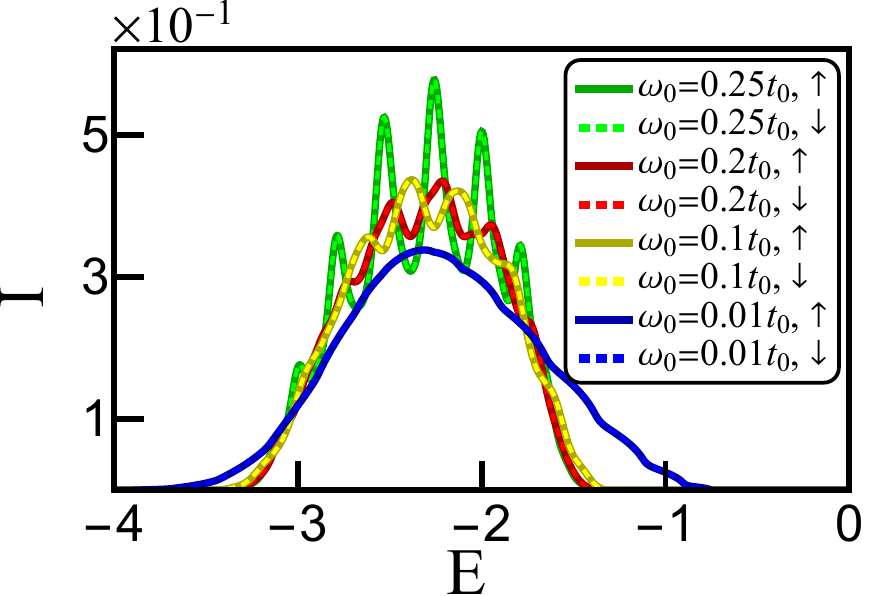}
\put(15,15){ \large\bfseries{(a)}}
\end{overpic}
\begin{overpic}[width=0.33\linewidth]{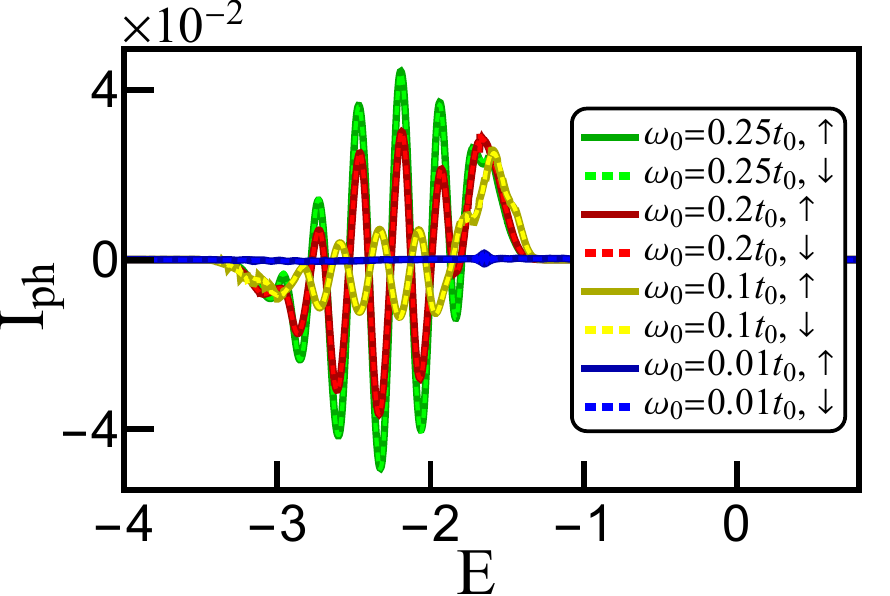}
\put(15,15){ \large\bfseries{(b)}}
\end{overpic}
\begin{overpic}[width=0.32\linewidth]{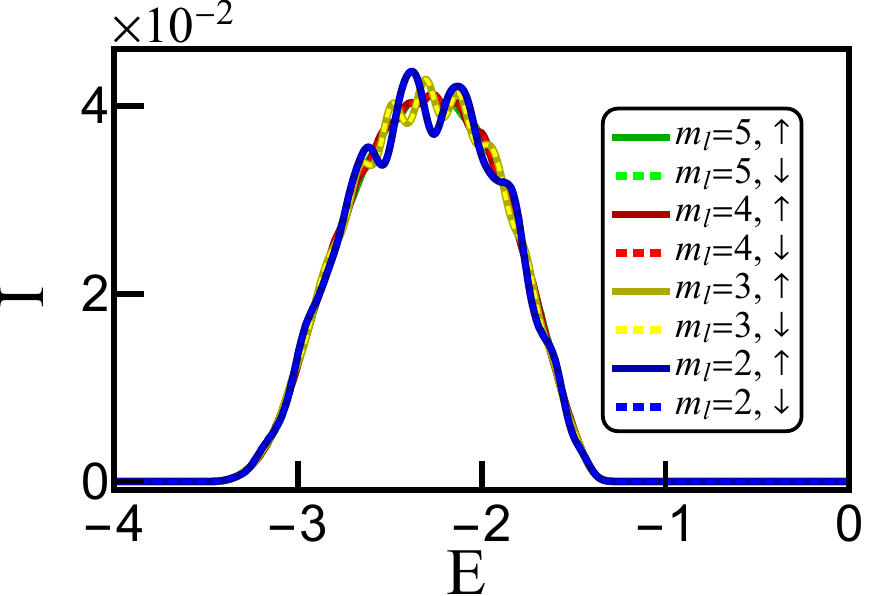}
\put(15,15){ \large\bfseries{(c)}}
\end{overpic}
\begin{overpic}[width=0.32\linewidth]{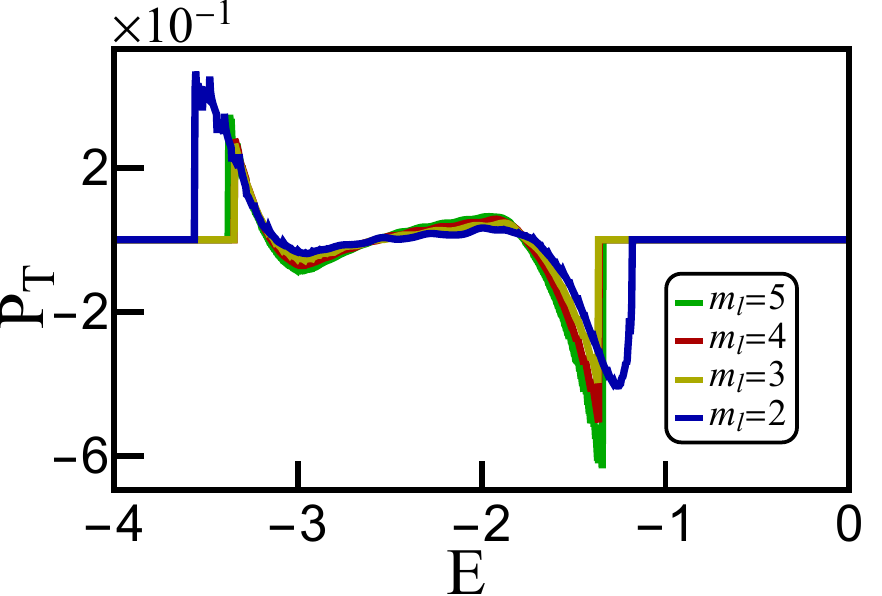}
\put(15,15){ \large\bfseries{(d)}}
\end{overpic}
\begin{overpic}[width=0.32\linewidth]{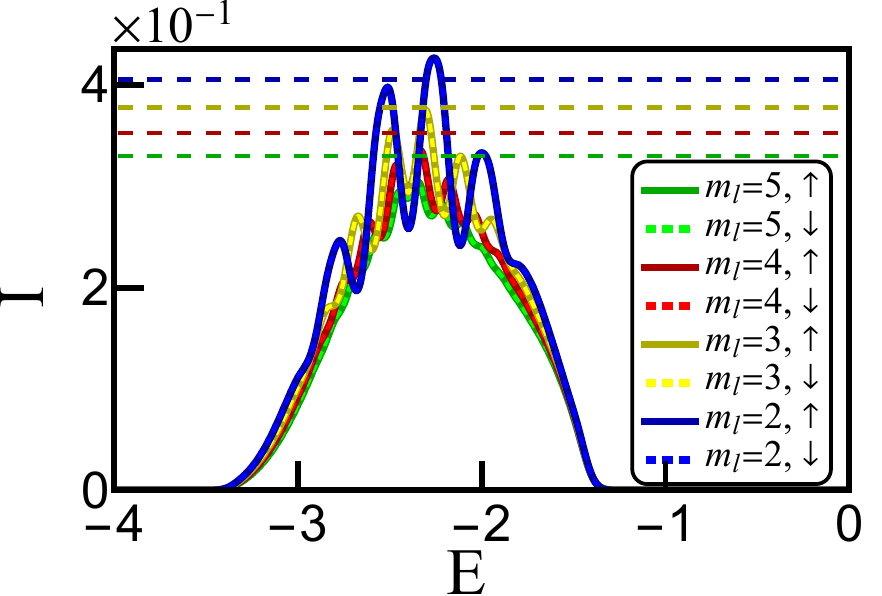}
\put(15,15){ \large\bfseries{(e)}}
\end{overpic}
\begin{overpic}[width=0.32\linewidth]{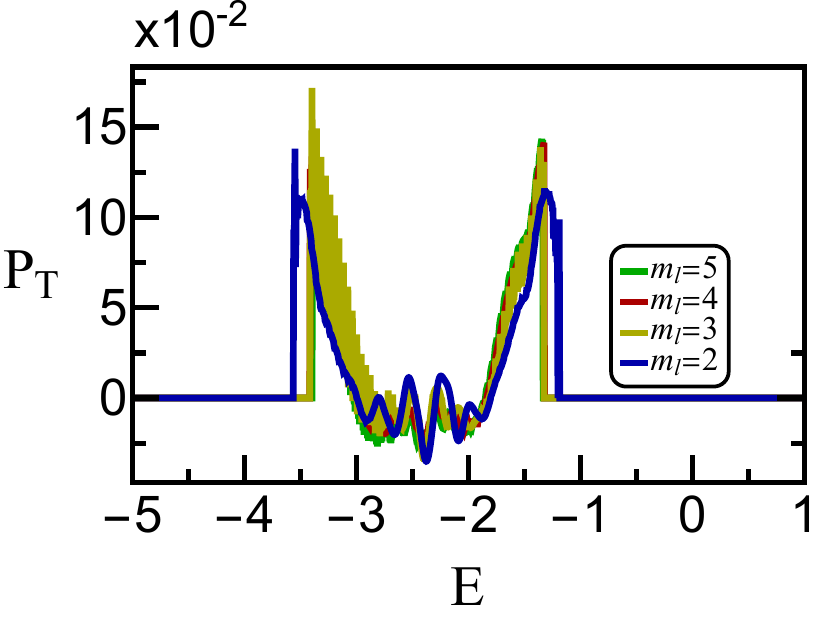}
\put(15,15){ \large\bfseries{(f)}}
\end{overpic}
\caption{\textbf{Energy-dependent figures with local e-ph interaction:}
\textbf{(a)} Variation of particle current coming out of the left bath for opposite lead magnetization and different phonon frequencies. 
\textbf{(b)} Energy-dependent phonon-bath current for different phonon frequencies. 
\textbf{(c)} Energy-dependent left-lead particle current for different system sizes. 
\textbf{(d)} Energy-dependent spin polarization for different system sizes. 
Current characteristics under the diagonal approximation described in subsection \ref{diagonal_approximation}: 
\textbf{(e)} Energy-dependent left-lead particle current for different system sizes within the diagonal approximation. The horizontal dashed lines show the corresponding energy-integrated particle current for each system size. 
\textbf{(f)} Energy-dependent spin polarization within the diagonal approximation. 
The default system size considered is $n_l=4, m_l=2$. 
The e-ph coupling follows the same structure as in \cite{Fransson_PhysRevB.102.235416}, with $\varepsilon_1=0$. 
The default parameter values, in units of $0.1$ eV, are 
$t_0=0.4,\varepsilon_0=-6t_0, t_1=0,$ $\lambda_0=t_0/40,$ $\lambda_1=0.1\lambda_0$, $\Gamma_0=t_0/4,$ $\omega_0=0.1t_0,$ and $T=0.25$ $(300\,\text{K})$. 
For the mixed coupling used in the diagonal-approximation figures, $t_1=0.1t_0$. 
The chemical potentials are $\mu_L=V/2, \mu_R=-V/2$, with $V=15t_0$. 
}
\label{local_interactions_figs}
\end{figure*}

This observation is further confirmed by the corresponding polarization plot shown in Fig.~\ref{ideal_fig}(d), where the spin polarization remains quite small at all energies for the different frequencies considered. As discussed earlier in the subsection \ref{numerical_details}, these polarization plots are obtained while accounting for residual numerical noise. This tolerance leads to the appearance of sharp cutoffs observed. 

The current contribution arising specifically from e-ph coupling is shown in Fig.~\ref{ideal_fig}(e). We find that the inelastic current associated with e-ph interactions is negligibly small at low phonon frequencies when compared to the current through the real leads. However, this contribution increases with increasing phonon frequency, indicating that low-frequency phonons are absorbed less efficiently than higher-frequency modes. Although the resulting phonon-bath currents remain smaller than the real lead current, they are not negligible, consistent with expectations in the weak e-ph coupling regime.

\par We also observe a correlation between the energy dependence of the e-ph current and the polarization profile shown in Fig.~\ref{ideal_fig}(d), with larger phonon-bath currents generally corresponding to enhanced polarization. However, because the phonon currents exhibit oscillatory behavior enforced by the overall current-conservation condition, $I_L = I_R$, similar oscillations appear in the polarization curve. This indicates that the net polarization, obtained after averaging over energy, is expected to be significantly smaller than the magnitude suggested by the individual peaks. A detailed discussion of this effect is deferred to Sec.~\ref{Sec:Results:B}.
In the same figure ~\ref{ideal_fig}(e), we also display the numerical noise associated with these calculations, estimated by the sum of the left, right, and phonon-bath currents, which should vanish in the ideal limit. We find that this quantity remains negligibly small compared to the physical current signals, thereby confirming the numerical stability and reliability of our results.
\par Finally, by plotting the left-lead current for different system sizes in Fig.~\ref{ideal_fig}(f), we observe that the energy-dependent current profiles do not change significantly with system size, indicating quasi-ballistic transport in the setup. This suggests that the e-ph coupling primarily leads to spectral renormalization and phase loss, rather than momentum loss. Suspecting that the global nature of the e-ph interaction may be responsible for this behavior, we next study a more local model with on-site e-ph interactions. The results of this analysis are discussed in the following subsection.

\begin{figure*}[t]
\centering

\begin{overpic}[width=0.32\linewidth]{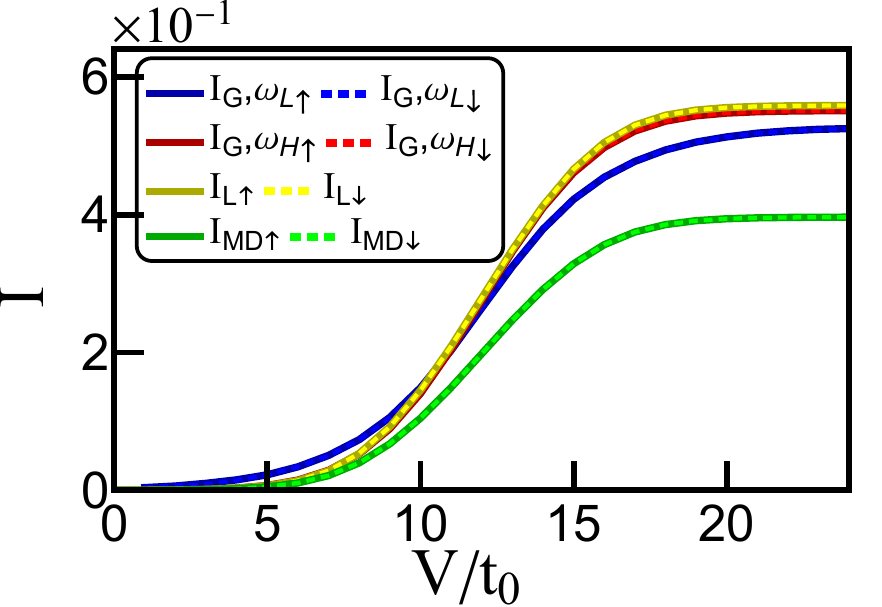}
\put(15,30){ \large\bfseries{(a)}}
\end{overpic}
\begin{overpic}[width=0.32\linewidth]{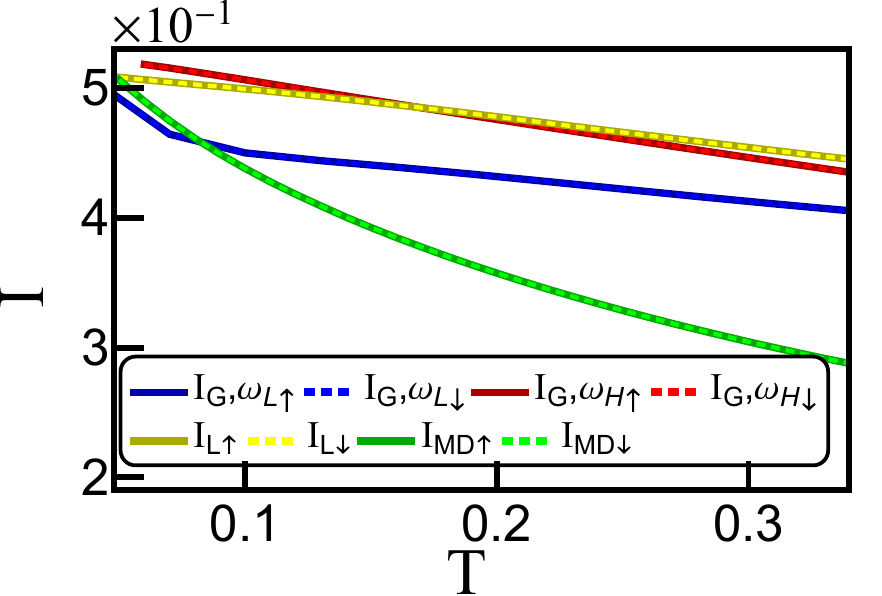}
\put(15,30){ \large\bfseries{(b)}}
\end{overpic}
\begin{overpic}[width=0.32\linewidth]{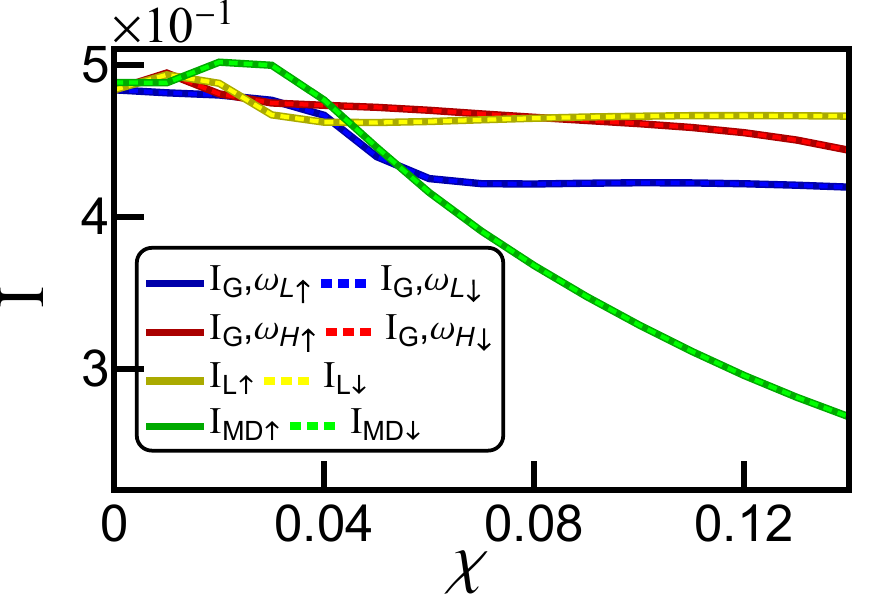}
\put(85,30){ \large\bfseries{(c)}}
\end{overpic}
\begin{overpic}[width=0.32\linewidth]{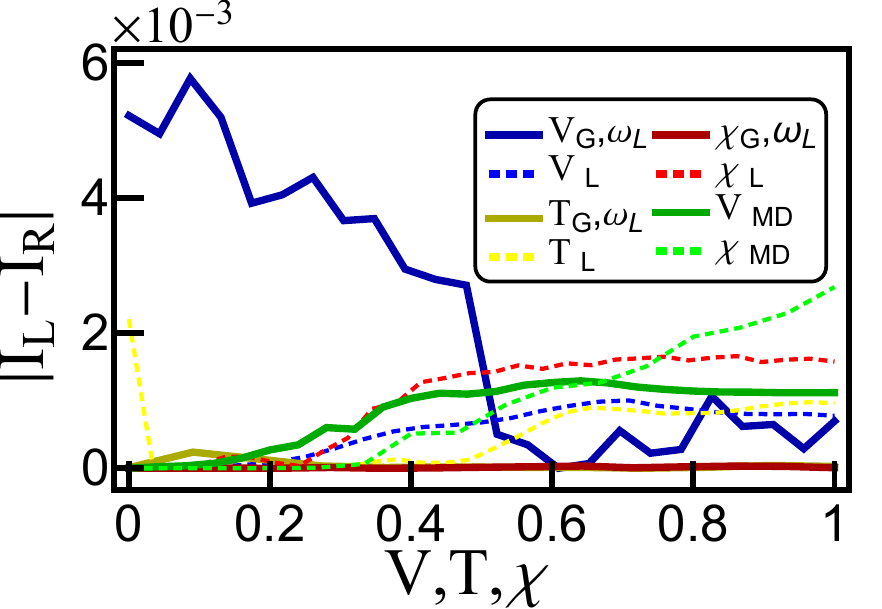}
\put(15,30){ \large\bfseries{(d)}}
\end{overpic}
\begin{overpic}[width=0.32\linewidth]{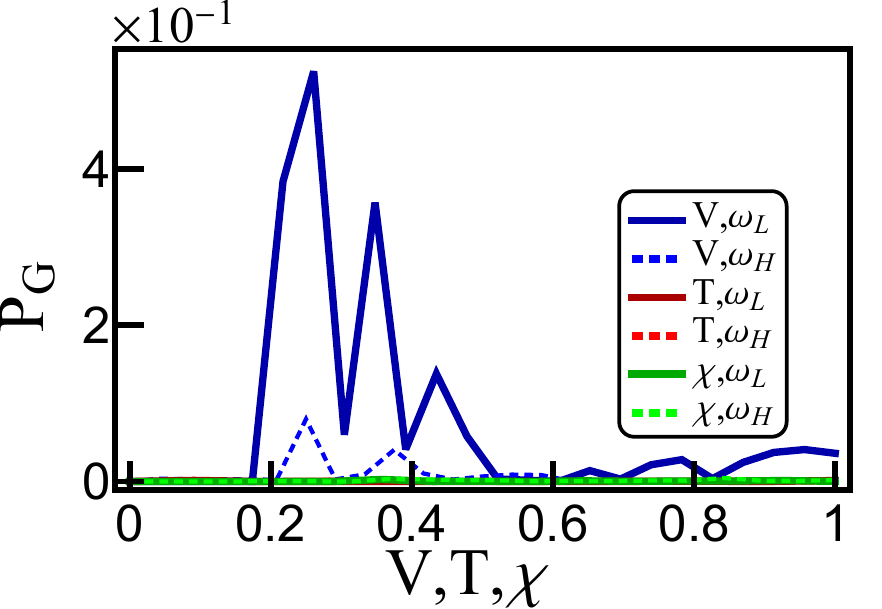}
\put(15,30){ \large\bfseries{(e)}}
\end{overpic}
\begin{overpic}[width=0.32\linewidth]{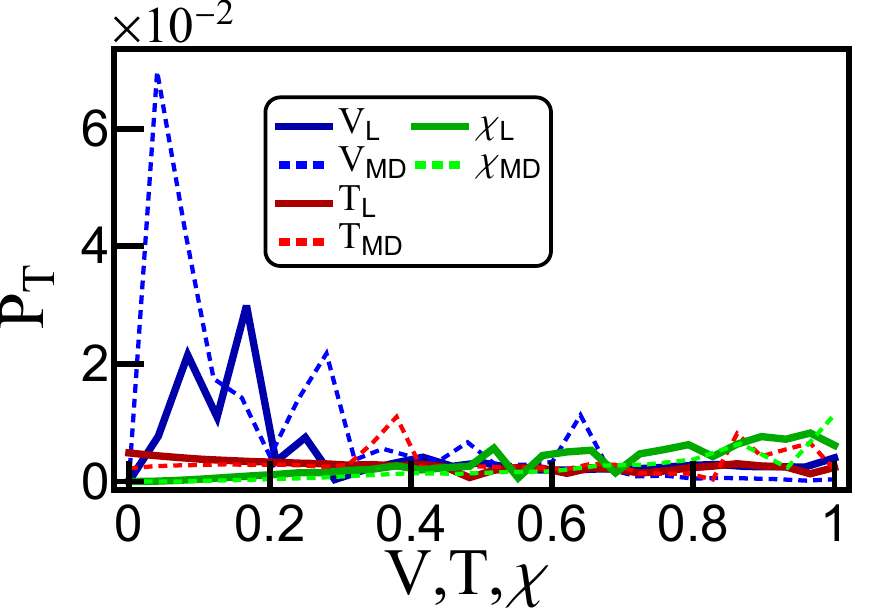}
\put(85,30){ \large\bfseries{(f)}}
\end{overpic}

\caption{\textbf{Energy integrated figures:} \textbf{(a)} Variation of particle current for global (G), local (L), and mixed-diagonal (MD) couplings with  voltage. For the global case, two different phonon frequencies are considered, `$\omega_0=\omega_L=0.01t_0, \omega_0=\omega_H=0.1t_0$', for the other two cases the phonon frequency is $\omega_0=0.1t_0$. Variation of particle current with \textbf{(b)} Temperature, and \textbf{(c)} e-ph coupling strength $\chi$.
\textbf{(d)} Difference between left and right bath current with normalised Voltage, Temperature, and  $\chi$ for p=0.5. The x-axis for the different parameters has been normalised according to their respective ranges, as shown in earlier graphs. polarization plots with normalised Voltage, Temperature, and e-ph interaction strength for \textbf{(e)} global coupling, \textbf{(f)} Local coupling, and mixed coupling with Diagonal Approximation. The default parameters for Global and Local figures are the same as stated in earlier figures. For Mixed-Diagonal, we use the system size $m_l=5, n_l=4$.}
\label{polarization_fig}
\end{figure*}

\subsection{Behavior with local interaction} \label{local_interaction}
We next examine how the transport behavior changes when a Holstein-like on-site e-ph interaction \cite{Galperin_2007} with $\varepsilon_1 \neq 0$ is introduced. For this case, we set the spin-independent hopping parameter to zero, $t_1 = 0$, while keeping the spin-dependent e-ph interaction finite, $\lambda_1 \neq 0$, and refer to this setup as the local (L) case.

In Fig.~\ref{local_interactions_figs}(a), we plot the current for different phonon frequencies. As in the global interaction case, the current spectrum becomes progressively smoother as the phonon frequency is reduced. Figure~\ref{local_interactions_figs}(b) further shows that, even for local coupling, phonon absorption increases with increasing phonon frequency.
The dependence of the current on system size is shown in Fig.~\ref{local_interactions_figs}(c). Here again, the energy-resolved current spectrum shows little variation with system size, indicating that the transport remains quasi-ballistic. This suggests that, even for local e-ph interactions, momentum relaxation is weak and the dominant effect of the coupling is spectral renormalization and phase loss. A similar behavior has been reported for a dephasing model previously \cite{datta_dephasing_and_momentum_PhysRevB.75.081301}, where a self-consistent treatment using the full corrective matrices leads to only phase loss while preserving momentum conservation.
Despite the presence of a non-negligible phonon bath contribution to the electron current in all the cases discussed above, we find that the particle currents corresponding to opposite lead magnetizations remain nearly indistinguishable. This indicates that weak e-ph coupling, even in the presence of local Holstein interactions and spin-dependent hopping, is insufficient to induce a significant CISS effect in chiral molecules. This conclusion is further supported by the polarization plotted for different system sizes in Fig.~\ref{local_interactions_figs}(d). 
\par Suspecting that the quasi-ballistic transport behavior observed in global and local cases is the reason for small spin polarization, we implement a diagonal approximation, similar to that employed in Ref.~\cite{datta_dephasing_and_momentum_PhysRevB.75.081301}, in the following subsection.

\subsection{Mixed coupling with diagonal approximation} \label{diagonal_approximation}
To explore the impact of enhanced dephasing on transport, we now adopt a diagonal approximation for the electron-phonon self-energies, keeping only the diagonal elements of the lesser and greater Green’s functions in Eq.~\eqref{self_consi_4}. The resulting e-ph-dependent matrices are then used in the retarded Green’s function of Eq.~\eqref{slef_consitent_1}, again retaining only the diagonal contributions. This approach makes the setup analogous to Büttiker voltage probes \cite{buttiker1986role,datta_dephasing_and_momentum_PhysRevB.75.081301}, effectively introducing phase-breaking as well as momentum breaking processes, and reducing the electronic mean free path. As a result, the system is expected to leave the quasi-ballistic regime and enter a super-diffusive transport regime.
For this case, we consider a generalized electron-phonon coupling where all the e-ph coupling parameters, $\varepsilon_1$, $t_1$, and $\lambda_1$, are nonzero.
\par In Fig.~\ref{local_interactions_figs}(e), we plot the current profile for different system sizes. In this case, the current decreases with increasing system size, although the reduction does not follow an exact $1/N$ scaling, indicating that the system is in a regime between ballistic and diffusive transport, i.e., the super-diffusive regime. The corresponding polarization, shown in Fig.~\ref{local_interactions_figs}(f), increases slightly with system size, but the enhancement is still not sufficient to produce large spin polarizations. In the next section, we examine the energy-integrated behavior for all the cases considered till now.

\section{Energy integrated behavior} \label{Sec:Results:B}

Finally, we study the energy-integrated total currents for the three types of interactions discussed in the previous section: the Fransson-like global e-ph interaction (G), the Holstein-like local interaction with spin-dependent nearest-neighbor hopping (L), and the mixed coupling with the diagonal approximation (MD). This energy-integrated current corresponds to the quantity typically measured in CISS experiments \cite{Fransson_What_it_CISS_https://doi.org/10.1002/ijch.202200046}
 and is directly related to the magnitude of the CISS effect.

We begin by plotting the energy-integrated currents as a function of voltage in Fig.~\ref{polarization_fig}(a). As expected, the particle currents are zero in the absence of a chemical potential difference between the leads. With increasing voltage, the currents rise non-monotonically and eventually saturate at higher voltages. For all interaction models, the particle currents under opposite lead magnetizations remain visually indistinguishable. This indicates that, irrespective of the interaction type or the amount of phonon absorption, the CISS effect remains negligible in the weak e-ph coupling regime.

Figure~\ref{polarization_fig}(b) shows the currents as a function of temperature. As anticipated, the currents decrease with increasing temperature due to enhanced thermal fluctuations \cite{Dubi_Temp_10.1063/5.0147886}. Examining the currents as a function of the e-ph coupling strength parameter $\chi$ in Fig.~\ref{polarization_fig}(c), we observe a general decrease, consistent with dissipation induced by interactions. Interestingly, for very weak coupling and relatively larger frequency ($\omega_0 = 0.1 t_0$), the currents show a modest increase for all the types of coupling. This counterintuitive enhancement, where weakly opening the inelastic channel increases the current, has been reported previously \cite{e-ph_PhysRevB.72.201101}.

\begin{figure*}[t]
\centering
\begin{overpic}[width=0.32\linewidth]{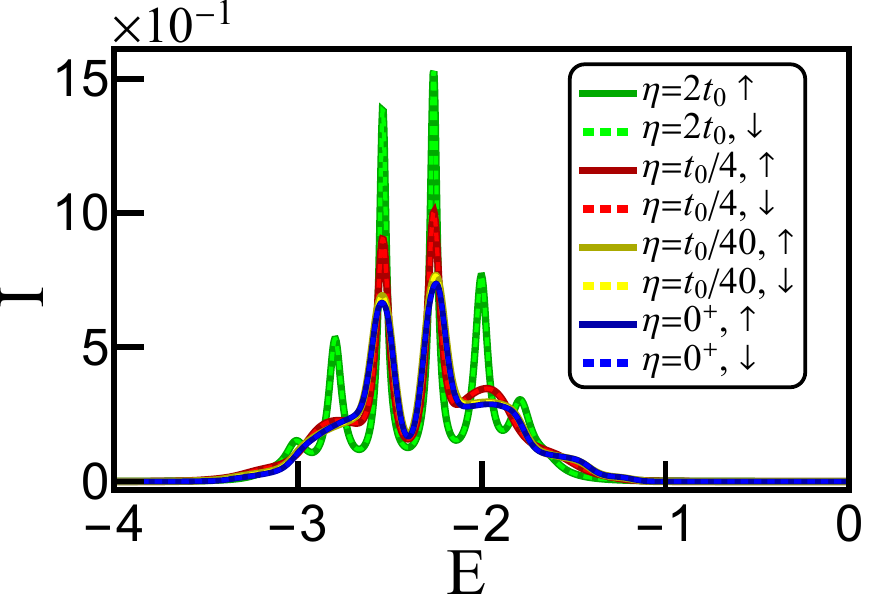}
\put(80,18){ \large\bfseries{(a)}}
\end{overpic}
\begin{overpic}[width=0.32\linewidth]{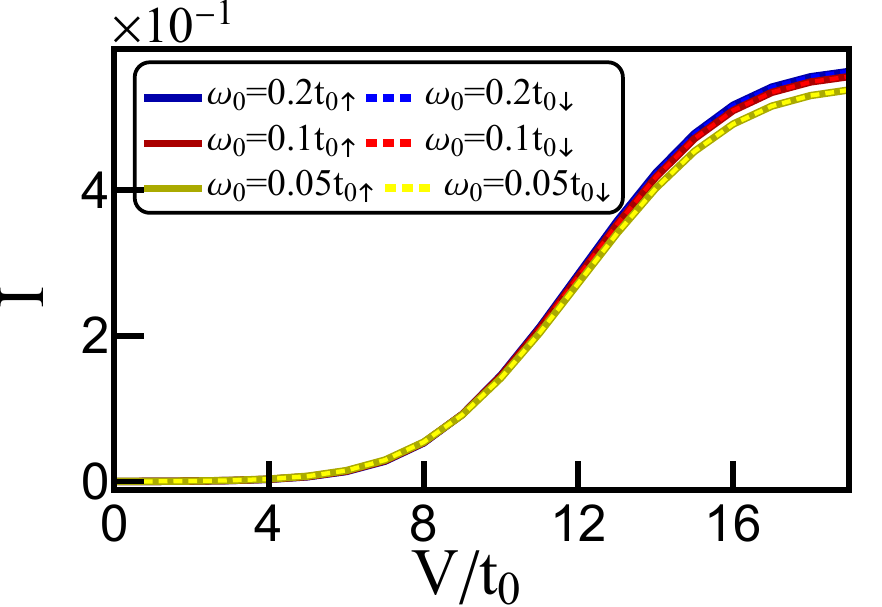}
\put(80,18){ \large\bfseries{(b)}}
\end{overpic}
\begin{overpic}[width=0.32\linewidth]{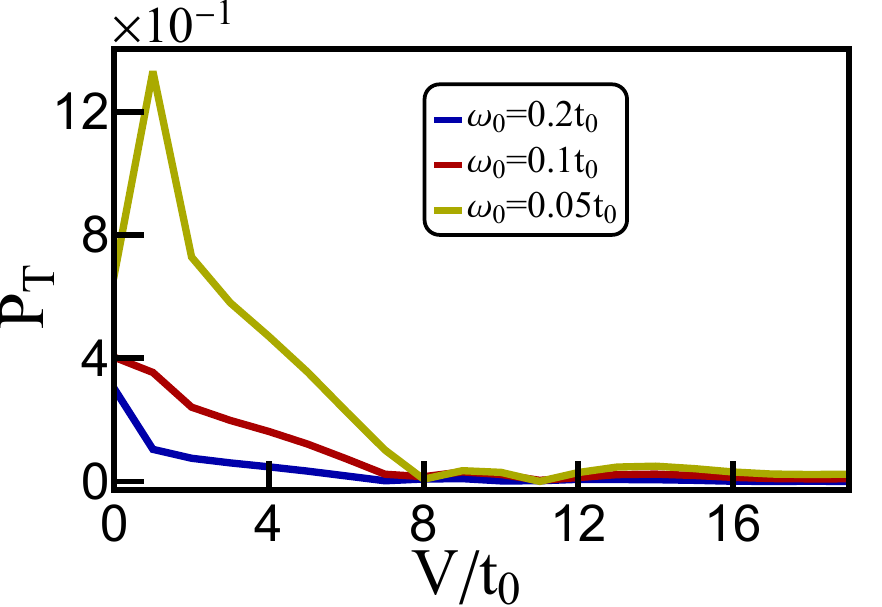}
\put(80,18){ \large\bfseries{(c)}}
\end{overpic}
\caption{ \textbf{Finite phonon linewidth figures:}
\textbf{(a)} Energy-dependent particle current flowing out of the left lead for opposite lead magnetizations and for different phonon linewidths $\eta$. The limit $\eta\to 0^+$ corresponds to an infinitely long-lived (sharp) phonon mode.
\textbf{(b)} Energy-integrated particle current flowing out of the left lead as a function of bias voltage for opposite lead magnetizations and for different phonon linewidths $\eta$.
\textbf{(c)} Spin polarization as a function of bias voltage for different phonon linewidths $\eta$.
Unless otherwise specified, all remaining parameters are the same as in the global-coupling case, with $\omega_0=0.1t_0$ and the reference linewidth $\eta=t_0/4$.}
\label{Damped_phonon_fig}
\end{figure*}

Based on these results, even at the energy-integrated level, the currents for opposite lead magnetizations remain nearly identical across all parameter regimes. This reinforces the conclusion that the present model does not produce significant spin polarization under weak e-ph coupling. To further validate our numerics, we verify current conservation by plotting the difference between the left and right bath currents in Fig.~\ref{polarization_fig}(d). The x-axis is normalized for all parameters between 0 and 1. We find that the difference remains negligibly small across the entire parameter range, confirming that the left and right currents are equal within numerical tolerance, a nontrivial consistency check of our implementation.

We also plot the corresponding  { absolute magnitude} of spin polarization for all parameters ($V$, $T$, $\chi$). In Fig.~\ref{polarization_fig}(e), the global interaction results show very small polarization. Notably, the energy-integrated polarization is smaller for higher-frequency phonons, despite their larger phonon currents, indicating that the correlation between phonon currents and polarization seen in energy-resolved results does not persist after integration. Figure~\ref{polarization_fig}(f) shows the polarization for the local and mixed-diagonal cases, which remain similarly negligible. Overall, even the integrated currents exhibit minimal spin polarization.

All our results indicate that, within SCBA and weak electron-phonon coupling, the molecular model considered here is insufficient to produce observable CISS effects. The lack of significant spin polarization suggests that additional physical ingredients or more advanced treatments beyond SCBA may be necessary to capture the essential mechanisms underlying chiral-induced spin selectivity.

{  \subsection{Effects of a finite phonon lifetime}

In all results presented so far, the vibrational mode has been treated as a sharp mode of frequency $\omega_0$.
Within the SCBA equations used in the main text, this corresponds to phonon-assisted processes that connect
electronic Green's functions at energies $E\pm\omega_0$ [cf. Eqs.~(8) and (11)], i.e.\ to a phonon spectral
function consisting of sharp peaks at $\pm\omega_0$.
While the phonon occupation is assumed to remain thermal, $n_B(\omega_0,T)$, this sharp-mode treatment
does not include an explicit finite phonon lifetime.

A recent study~\cite{fransson2025chiralinducedspinpolarized} argued that vibrational relaxation, arising from coupling of the molecular vibration
to an external thermal bath, may be important for generating a sizable CISS signal. To model this effect at a
phenomenological level, we replace the sharp phonon peaks by Lorentzians of finite width $\eta$, which mimics
a finite vibrational lifetime. This is the same spirit as the Lorentzian broadening introduced in Appendix~B,
where $\delta(x)\rightarrow \delta_\eta(x)$ with
\begin{equation}
\delta_\eta(x)=\frac{1}{\pi}\frac{\eta}{x^2+\eta^2},
\qquad \eta>0 .
\end{equation}

Accordingly, we describe the broadened phonon spectrum by
\begin{equation}
L(\omega-\omega_0,\eta)
=
\frac{1}{\pi}\frac{\eta}{(\omega-\omega_0)^2+\eta^2},
\end{equation}
and define
$L_-^\eta\equiv L(\omega-\omega_0,\eta)$ and
$L_+^\eta \equiv L(\omega+\omega_0,\eta),$
which are centered at $+\omega_0$ and $-\omega_0$, respectively.
Keeping the phonon bath in thermal equilibrium, with $n_0\equiv n_B(\omega_0,T)$, the sharp-mode SCBA
expressions are generalized to

\begin{equation}
\begin{split}
\Gamma_{\rm ph}(E)
&=
i\,M \Bigg[
\int d\omega\,
\big((1+n_0)L_-^\eta+n_0L_+^\eta\big)
G^{>}(E-\omega)
\\
&
-
\int d\omega\,
\big(n_0L_-^\eta+(1+n_0)L_+^\eta\big)
G^{<}(E-\omega)
\Bigg] M ,
\end{split}
\label{Gamma_ph_damp}
\end{equation} 
while the corresponding phonon-induced in-scattering self-energy becomes
\begin{align}\label{phonon_inscattering_damped}
\Sigma_{\mathrm{ph}}^<(E)
&=
M
\int d\omega
\Big[
n_0 L_-^\eta
+
(1+n_0)L_+^\eta
\Big]
G^{<}(E-\omega)
M 
\end{align}
In the limit $\eta\to 0^+$, one recovers the sharp-mode SCBA expressions used in the main text.
We note that this procedure is a phenomenological way to include a finite phonon lifetime while  
assuming thermal phonon occupations. It is therefore not a fully self-consistent phonon treatment, but it provides a useful benchmark for assessing whether finite vibrational linewidths can qualitatively alter the current and polarization.

We now analyze the current and spin polarization obtained for several values of $\eta$, corresponding to different
strengths of vibrational relaxation, and compare them with the sharp-mode results discussed above. 
In Fig.~\ref{Damped_phonon_fig}(a), we plot the energy-dependent particle current for different values of the phonon linewidth $\eta$. We find that the detailed current profile changes noticeably as $\eta$ is varied, with the number, position, and sharpness of the peaks being modified by the finite phonon linewidth. Since $\eta$ enters the phonon-assisted self-energies through a Lorentzian broadening of the phonon spectrum, changing $\eta$ alters the energy averaging in the self-consistent electronic self-energies and can therefore substantially reshape the energy-resolved current. This comparison is mainly intended to show that, although the finite phonon linewidth can significantly modify the detailed energy-resolved current profile, these changes do not translate into any substantial enhancement of the transport spin polarization.

In Fig.~\ref{Damped_phonon_fig}(c), the energy-integrated particle current shows a non-monotonic dependence on the applied bias, with the saturation current exhibiting only a moderate dependence on the phonon linewidth. Importantly, as seen in Fig.~\ref{Damped_phonon_fig}(c), the transport spin polarization remains small for all values of $\eta$ considered, indicating that the inclusion of a finite phonon linewidth does not qualitatively alter our central conclusion.}

\section{Conclusion}\label{sec:conclusion}
We investigated spin polarization in electron transport through a helical molecular junction coupled to vibrational modes, focusing on a widely used vibrational mechanism proposed to enhance CISS~\cite{Fransson_PhysRevB.102.235416}. In contrast to previous studies, we solved the NEGF equations fully self-consistently within the self-consistent Born approximation (SCBA), thereby avoiding uncontrolled evaluation of the electron-phonon self-energies. Within this weak-coupling framework, the resulting steady-state spin polarization is strongly suppressed and remains negligible over a wide range of parameters. Our results therefore, constrain vibrationally assisted CISS in two-terminal, single-orbital junction models in the weak electron-phonon coupling regime.

We performed the analysis for three electron-phonon coupling schemes: global, local, and a mixed scheme employing a diagonal self-energy approximation. For both the global and local couplings, electron-phonon interactions primarily renormalize the electronic spectrum, and the current exhibits only a weak dependence on system size, consistent with predominantly quasi-ballistic transport. In contrast, introducing the diagonal approximation qualitatively alters the size dependence of the current, producing a pronounced decrease with increasing system size. Importantly, across all three schemes, the energy-integrated spin polarization remains very small. Although the energy-resolved calculations can show localized polarization features (in some cases correlated with phonon-bath currents), these features are strongly suppressed upon energy integration.
{ We further tested a phenomenological finite-phonon-linewidth extension, motivated by recent proposals emphasizing vibrational relaxation, and found that although a finite linewidth can substantially reshape the energy-resolved current, it does not produce any significant enhancement of the energy-integrated transport spin polarization within the present self-consistent treatment.
}

Since the weak-coupling, two-terminal models studied here do not yield a sizable steady-state polarization even when the transport regime changes under approximations, it is natural to seek additional mechanisms. Possible missing ingredients include multi-orbital structure~\cite{Fransson_What_it_CISS_https://doi.org/10.1002/ijch.202200046,Aharony_Ciss_PhysRevB.102.035445}, stronger-coupling effects, electron--electron interactions, or explicitly nonequilibrium spin-dependent environments.

\section{Acknowledgement}
We thank Jonas Fransson for his comments on the manuscript and thank Lev Kantorovich for his suggestion of using Pulay method for numerical convergence. This research was supported by the ISF Grants No.~3105/23.
\section{DATA AVAILABILITY}
The code supporting the findings of this study are
openly available at \cite{github_codes}.

\appendix
\begin{widetext}
\section{Derivation of the NEGF equations using the Quantum Langevin Equation Approach}\label{deriv_eqns}
We try to derive the relevant equation of motion for our study using the so called Quantum Langevin equation method pioneered by Dhar et. al\cite{Dhar_PhysRevB.73.085119}.  Specifically, major part of this derivation is similar to the one presented in the book \cite{Datta_2005}. To start, we consider a molecular junction consisting of a molecule that is coupled to two electronic reservoirs (left and right 
leads), while interacting with vibrational degrees of freedom (phonons).
Throughout we set $\hbar=1, k_B=1$ and adopt the Heisenberg picture where the
time-evolution of an operator $\hat{O}$ is governed by,
\begin{equation}
\frac{d\hat{O}}{dt} = i[H_{\mathrm{tot}},\hat{O}].
\end{equation}
The total Hamiltonian is written as a sum of five contributions,
\begin{equation}
H_{\text{tot}} = H_{\text{el}} + H_{\text{ph}} + H_{\text{el-ph}}
               + H_{\text{leads}} + H_{\text{e-lead}} ,
\end{equation}
each of which is defined explicitly below. The molecule is modeled using a tight-binding representation,
\begin{equation}
H_{\text{el}} = \sum_{i,j}
H_{ij}\,c_i^\dagger c_j ,
\end{equation}
where $c_i^\dagger$ ($c_i$) creates (annihilates) an electron on molecular site
$i$, and $H_{ij}$ describes onsite energies and coherent hopping amplitudes
within the molecular structure. 
The molecular vibrations are described as independent harmonic oscillators,
\begin{equation}
H_{\text{ph}} = \sum_{\beta}
\omega_\beta \left( b_\beta^\dagger b_\beta + \tfrac{1}{2} \right),
\end{equation}
where $b_\beta^\dagger$ ($b_\beta$) creates (annihilates) a phonon in mode
$\beta$ of frequency $\omega_\beta$.
We consider linear coupling between the electronic density matrix and phonon
displacements,
\begin{equation}
H_{\text{el-ph}} =
\sum_{i,j,\beta} M_{ij}^\beta \,
c_i^\dagger c_j \big(b_\beta^\dagger + b_\beta\big),
\end{equation}
where $M_{ij}^\beta$ denotes the coupling matrix elements. We assume
$M^\beta$ to be Hermitian, ensuring the overall Hamiltonian is Hermitian.
The electronic reservoirs are taken as non-interacting Fermi seas in
thermodynamic equilibrium at their respective chemical potentials $\mu_L$ and
$\mu_R$,
\begin{equation}
H_{\text{leads}} =
\sum_{\alpha} \varepsilon^L_{\alpha}\, c_{\alpha}^\dagger c_{\alpha}
+\sum_{\alpha'} \varepsilon^R_{\alpha'}\, c_{\alpha'}^\dagger c_{\alpha'} ,
\end{equation}
with $c_{\alpha}$ ($c_{\alpha'}$) denoting annihilation operators in the left
(right) lead.
Electron tunneling between molecule and reservoirs is described by,
\begin{equation}
H_{\text{e-lead}}
=
\sum_{i,\alpha}
\Big( V^L_{i\alpha}\, c_i^\dagger c_\alpha + \mathrm{h.c.} \Big)
+
\sum_{i,\alpha'}
\Big( V^R_{i\alpha'}\, c_i^\dagger c_{\alpha'} + \mathrm{h.c.} \Big) .
\end{equation}

We now derive the operator equations of motion. Applying the Heisenberg
equation to the molecular annihilation operator yields
\begin{align}
\dot{c}_i(t)
&=-i\sum_j H_{ij} c_j(t)
-i\sum_{j\beta} M_{ij}^\beta c_j(t)
   \big( b_\beta^\dagger(t) + b_\beta(t) \big) \nonumber\\
&\quad
-i\sum_\alpha V^L_{i\alpha} c_\alpha(t)
-i\sum_{\alpha'} V^R_{i\alpha'} c_{\alpha'}(t).
\label{eq:c_eom}
\end{align}
It is useful to rewrite this compactly in vector notation as
\begin{equation} \label{first_c_eqn}
|\dot{c}(t)\rangle
=
-i H\,|c(t)\rangle
-i\sum_{\beta} M^\beta \big( |c(t)b_\beta\rangle + |c(t)b_\beta^\dagger\rangle \big)
-i V^L |c(t)\rangle_L
-i V^R |c(t)\rangle_R ,
\end{equation}
where the vectors encode the molecular, left-lead, and right-lead operators.

The corresponding equations for the lead fermions read
\begin{align}
\dot{c}_\alpha(t)
&= -i\varepsilon^L_\alpha c_\alpha(t)
   - i\sum_j V^{L*}_{j\alpha} c_j(t), \\
\dot{c}_{\alpha'}(t)
&= -i\varepsilon^R_{\alpha'} c_{\alpha'}(t)
   - i\sum_j V^{R*}_{j\alpha'} c_j(t),
\end{align}
while those for the phonon operators are
\begin{align}
\dot{b}_\beta(t)
&= -i\omega_\beta b_\beta(t)
   - i\sum_{ij} M^\beta_{ij} c_i^\dagger(t) c_j(t), \\
\dot{b}^\dagger_\beta(t)
&=i \omega_\beta b^\dagger_\beta(t)
   + i\sum_{ij} M^\beta_{ij} c_i^\dagger(t) c_j(t).
\end{align}

Equations \eqref{eq:c_eom} form the starting point for our
non–equilibrium quantum transport analysis. In the following, we formally solve
the reservoir and phonon equations and substitute the results into
\eqref{eq:c_eom}. This yields a closed integro-differential quantum Langevin
equation for the molecular operators, which can then be systematically mapped
to the standard NEGF formulation.
\subsection*{Solving the lead equation of motion (EOM)}

The left-lead operators satisfy the vector equation
\begin{equation}
|\dot{c}(t)\rangle_L = -i H^L |c(t)\rangle_L - i V^{L\dagger} |c(t)\rangle,
\end{equation}
where $H^L$ is the single-particle Hamiltonian of the left lead and $V^L$ the
molecule--left-lead coupling matrix. The homogeneous part gives the free
evolution,
\begin{equation}
|c^{(0)}(t)\rangle_L = e^{-i H^L (t-t_0)} |c(t_0)\rangle_L.
\end{equation}
The inhomogeneous equation can be solved using the integrating-factor method,
which yields
\begin{equation}
|c(t)\rangle_L = |c^{(0)}(t)\rangle_L
- i \int_{t_0}^{t} dt' \, e^{-i H^L (t-t')} V^{L\dagger} |c(t')\rangle.
\end{equation}
We identify the retarded Green's function (matrix) of the left lead as
\begin{equation}
g^L(t-t') \equiv -i\,\Theta(t-t')\,e^{-i H^L (t-t')}.
\end{equation}
An entirely analogous expression holds for the right lead,
\begin{equation}
|c(t)\rangle_R = |c^{(0)}(t)\rangle_R
- i \int_{t_0}^{t} dt' \, e^{-i H^R (t-t')} V^{R\dagger} |c(t')\rangle,
\end{equation}
with the corresponding retarded Green's function
$g^R(t-t')=-i\Theta(t-t')e^{-iH^R(t-t')}$.

To derive the equation of motion for the fermionic operators of the molecule, we
substitute these solutions into the system EOM and separate the lead
contribution into two parts:
\begin{enumerate}
    \item a noise term depending on the initial lead operators
          $|c(t_0)\rangle_{L,R}$,
    \item a damping term involving integrals over past system operators
          $|c(t')\rangle$.
\end{enumerate}
Following the standard quantum Langevin approach (see, e.g., \cite{Dhar_PhysRevB.73.085119}), we define
the \textbf{lead noise operators} as
\begin{equation}
|\eta_L(t)\rangle = iV^L g^L(t-t_0)\,|c(t_0)\rangle_L, 
\qquad
|\eta_R(t)\rangle = iV^R g^R(t-t_0)\,|c(t_0)\rangle_R.
\end{equation}
These terms account for random fluctuations originating from the leads in the
distant past ($t_0 \to -\infty$). Their ensemble averages vanish,
\[
\langle \eta_{L,R,\alpha}(t) \rangle = 0,
\]
while their correlations determine the statistical properties of the
reservoirs.

Next, we define the \textbf{lead self-energy kernels} as,
\begin{equation}
\Sigma^r_{L}(t-t') = V^L g^L(t-t') V^{L\dagger}, 
\qquad
\Sigma^r_{R}(t-t') = V^R g^R(t-t') V^{R\dagger}.
\end{equation}
Substituting everything into the equation of motion \eqref{eq:c_eom}, we obtain,
\begin{align} 
 |\dot{c}(t)\rangle
 &=-iH|c(t)\rangle- i\,|\eta_L(t)\rangle
       - i\int_{t_0}^{t} dt' \, \Sigma^r_{L}(t-t') |c(t')\rangle  - i\,|\eta_R(t)\rangle
       - i\int_{t_0}^{t} dt' \, \Sigma^r_{R}(t-t') |c(t')\rangle
  \nonumber\\&\quad + \text{Phonon terms}.
\label{second_eom}
\end{align}
The above equation is a quantum Langevin equation for the system
fermions, with explicit noise terms $|\eta_{L,R}\rangle$ and memory kernels
$\Sigma^r_{L,R}$ describing dissipation due to the real leads. Once we know these contribution we can directly write the steady state inscattering and broadening matrices for an energy `E' due to the real leads, using the relations
\cite{Dhar_PhysRevB.73.085119},
\begin{align}
\Sigma^<_{L,ij}(E)=i\langle \tilde{\eta}^\dagger_{L,j}(E) \tilde{\eta}_{L,i}(E) \rangle 
\end{align}
where, $\tilde{\eta}_{L,j}(E)$ is the $j^{th}$ component of the Fourier transform of the noise vector. Here, $\Sigma^<_{L}(E)=if_L(E)\Gamma_L$, as the left lead has well defined chemical potential.
Similarly we can also find the broadening matrix using the retarded kernel,
\begin{align}
\Gamma_L(E)=i(\tilde{\Sigma}^r_{L}(E)-\tilde{\Sigma}^a_{L}(E))
\end{align}
Again a function of $\tilde{\Sigma}^r_{L}(E)$ signifies Fourier transform of the damping kernel, $\tilde{\Sigma}^a_{L}(E)=\tilde{\Sigma}^{r\dagger}_{L}(E)$.
Once we have both the contribution, we can find the particle current at an energy `E', using the equations presented in the section \ref{sec:NEGF_Equations}. We are assuming a flat spectrum so, $\Gamma_L$ is a constant diagonal matrix with non-zero elements only at first and second diagonal, these element are proportional to the square of system-lead coupling. Similar relations hold for the right lead terms. In the following section, we try to write the contributions due to e-ph coupling.
\subsection*{Solving the electron--phonon equation of motion (EOM)}

We now analyze the dynamics of composite operators involving one electron and
one phonon. This will allow us to identify the phonon-induced noise and
damping acting on the electronic system.
We define,
\begin{equation}
C^{ph}_{n\gamma}(t) \equiv c_n(t)\, b_\gamma(t),
\end{equation}
and similarly, later, $K^{ph}_{n\gamma}(t)\equiv c_n(t)\,b^\dagger_\gamma(t)$.
The time derivative of $C^{ph}_{n\gamma}$ follows from the product rule:
\begin{equation}
\frac{d}{dt}\big(c_n b_\gamma\big)
= \dot c_n\, b_\gamma + c_n\,\dot b_\gamma.
\end{equation}
Using the equations of motion for $c_n$ and $b_\gamma$ we obtain
\begin{align}
\frac{d}{dt}\big(c_n b_\gamma\big)
&=\left(-i\sum_{j}H_{nj}c_j
        -i\sum_{j\beta}M^\beta_{nj}c_j(b_\beta^\dagger + b_\beta)
        -i\sum_{\alpha}V^L_{n\alpha} c_\alpha
        -i\sum_{\alpha'}V^R_{n\alpha'} c_{\alpha'}\right) b_\gamma \nonumber\\
&\quad + c_n\left(-i\omega_\gamma b_\gamma
                 -i\sum_{ij}M^{\gamma}_{ij}c_i^\dagger c_j \right).
\label{eq:dCdt_raw}
\end{align}
 We neglect direct lead--phonon terms, `$c_\alpha b_\gamma$', since phonons couple only
      to the molecular fermions and not lead fermions, and we expect their correlation to go to zero
      $\langle c_\alpha b_\gamma\rangle\approx 0$.
Using these approximations, we separate the free part of
\eqref{eq:dCdt_raw} and move it to the left-hand side:
\begin{align}
\frac{d}{dt} C^{ph}_{n\gamma}(t)
+ i\omega_\gamma C^{ph}_{n\gamma}(t)
+ i\sum_{j}H_{nj} C^{ph}_{j\gamma}(t)
&=-i\sum_{j\beta}M^\beta_{nj}c_j(b_\beta^\dagger + b_\beta )b_\gamma -i\sum_{ij}M^{\gamma}_{ij}c_nc_i^\dagger c_j 
\end{align}
We assume that the phonon cloud is at equilibrium at temperature $T$, so $\langle b_\beta b_\gamma\rangle=0$.
Now, simplifying by taking expectation values in place of product of multiplication of operator. This is similar to a mean field approximation,
\begin{align}
\frac{d}{dt} C^{ph}_{n\gamma}(t)
+ i\omega_\gamma C^{ph}_{n\gamma}(t)
+ i\sum_{j}H_{nj} C^{ph}_{j\gamma}(t)
&=-i\sum_{ij\beta}M^\beta_{ij} \langle b_\beta^\dagger b_\gamma \rangle \delta_{ni} c_j  -i\sum_{ij\beta}M^{\beta}_{ij}\langle c_nc_i^\dagger\rangle \delta_{\beta \gamma}  c_j 
\end{align}
Using the commutation and anitcommuation relation for Boson and Fermions respectively and substituting for the delta functions, we get,
\begin{align}
\frac{d}{dt} C^{ph}_{n\gamma}(t)
+ i\omega_\gamma C^{ph}_{n\gamma}(t)
+ i\sum_{j}H_{nj} C^{ph}_{j\gamma}(t)
&= -i\sum_{ij\beta}M^{\beta}_{ij}
   \Big(
     \langle c_i^\dagger(t) c_n(t)\rangle
     \langle b_\beta^\dagger(t) b_\gamma(t)\rangle \nonumber\\
&\qquad\qquad\qquad\quad
   +\langle c_n(t) c_i^\dagger(t)\rangle
    \langle b_\gamma(t) b_\beta^\dagger(t)\rangle
   \Big) c_j(t).
\label{eq:C_scalar}
\end{align}
Using the  retarded propagator for the composite operator,
\begin{equation}
g_{1,\gamma}^{ph}(t-t') \equiv -i\,\Theta(t-t')\,e^{-i (\omega_\gamma+H) (t-t')}.
\end{equation}
The formal solution of the above equation reads,
\begin{align}
C_{n\gamma}^{ph}(t)
&= C^{0\,ph}_{n\gamma}(t) \nonumber\\
&\quad - i \int_{t_0}^{t} dt' \, 
\sum_{ij\beta}M^{\beta}_{ij}
   g_{1,n\gamma}^{ph}(t-t')\Big(
     \langle c_i^\dagger(t') c_n(t')\rangle
     \langle b_\beta^\dagger(t') b_\gamma(t')\rangle  +\langle c_n(t') c_i^\dagger(t')\rangle
    \langle b_\gamma(t') b_\beta^\dagger(t')\rangle
   \Big) c_j(t').\label{eq:C_general}
\end{align}
Taking the lowest order approximation in the electron phonon interaction matrix $M^\beta$, we can take, $g_{1,n\gamma}^{ph}(t-t')c_n(t')b_\gamma(t')\approx c_n(t)b_\gamma(t)$ inside the integral. So,

\begin{align}
C_{n\gamma}^{ph}(t)
&\approx C^{0\,ph}_{n\gamma}(t) \nonumber\\
&\quad - i \int_{t_0}^{t} dt'\Theta(t-t') \, 
\sum_{ij\beta}M^{\beta}_{ij}
   \Big(
     \langle c_i^\dagger(t') c_n(t)\rangle
     \langle b_\beta^\dagger(t') b_\gamma(t)\rangle  +\langle c_n(t) c_i^\dagger(t')\rangle
    \langle b_\gamma(t) b_\beta^\dagger(t')\rangle
   \Big) c_j(t')+\mathcal{O}[(M^\beta)^2]\label{eq:C_general}
\end{align}

We now introduce the electronic correlation matrices,
\begin{align}
G^N_{i,j}(t,t') &\equiv \langle c_j^\dagger(t') c_i(t)\rangle, &
G^P_{i,j}(t,t') &\equiv \langle c_i(t) c_j^\dagger(t')\rangle,
\end{align}
\begin{align}
B^{ab}_{\gamma\beta}(t,t') &\equiv \langle b_\beta^\dagger(t') b_\gamma(t)\rangle,\\
B^{em}_{\gamma\beta}(t,t') &\equiv \langle b_\gamma(t) b_\beta^\dagger(t')\rangle.
\end{align}
In thermal equilibrium at temperature $T$, these become (in the mode basis)
\begin{align}
B^{ab}_{\gamma\beta}(t,t')
&=\delta_{\gamma\beta} e^{-i\omega_\gamma (t-t')}\, n_B(\omega_\gamma,T)\, \\
B^{em}_{\gamma\beta}(t,t')
&=\delta_{\gamma\beta} e^{-i\omega_\gamma (t-t')}\, \big(1+n_B(\omega_\gamma,T)\big)\, 
\end{align}
with the Bose function
\begin{equation}
n_B(\omega,T) = \frac{1}{e^{\omega/T} - 1}.
\end{equation}

In the equilibrium phonon basis, the sum over $\beta$ collapses due to
$B_{\gamma\beta}\propto\delta_{\gamma\beta}$. Writing this explicitly we obtain,
\begin{align}
|C_\gamma^{ph}(t)\rangle
&= |C^{0\,ph}_{\gamma}(t)\rangle \nonumber\\
&\quad - i  \int_{t_0}^{t} dt' \Theta(t-t')\,e^{-i\omega_\gamma (t-t')}
\Big( n_B(\omega_\gamma,T)\,  G^N(t,t') 
+\big(1+n_B(\omega_\gamma,T)\big)\, G^P(t,t')\Big)
M^\gamma\,|c(t')\rangle.
\label{eq:C_final}
\end{align}
A completely analogous derivation for the operator
$K^{ph}_{n\gamma}(t)\equiv c_n(t)\,b^\dagger_\gamma(t)$ gives
\begin{align}
|K_{\gamma}^{ph}(t)\rangle
&= |K^{0\,ph}_{\gamma}(t)\rangle \nonumber\\
&\quad - i  \int_{t_0}^{t} dt'\Theta(t-t') \,e^{i\omega_\gamma (t-t')}
\Big( n_B(\omega_\gamma,T)\,  G^P(t,t') 
    +\big(1+n_B(\omega_\gamma,T)\big)\, G^N(t,t')\Big)
M^\gamma\,|c(t')\rangle.
\label{eq:K_final}
\end{align}
Together, Eqs.~\eqref{eq:C_final} and~\eqref{eq:K_final} encode the influence
of the phonon bath on the electronic system in terms of equilibrium Bose
factors, electronic correlation functions, and the composite propagator
$g_{1,\gamma}^{ph}$.

Now we define the noise and damping term due to the electron phonon coupling. First the noise terms,

\begin{equation} \label{phonon_noise}
|\eta_\gamma^{ph}(t,t_0)\rangle = M^\gamma \left(g^{ph}_{1,\gamma}(t-t')|{C}_\gamma^{ph}(t_0)\rangle+g^{ph}_{2,\gamma}(t-t_0)|{K}_\gamma^{ph}(t_0)\rangle\right)
\end{equation}
where `$g^{ph}_{2,\gamma}(t-t_0)$' is the retarded propogator for $|{K}_\gamma^{ph}(t_0)\rangle$.  The damping term is given as,
\begin{align}
\Sigma_{\gamma}^{r,ph}(t-t') &=\Theta(t-t')\Big[ e^{-i\omega_\gamma (t-t')} M^\gamma \left(n_B(\omega_\gamma,T) G^N(t,t') +(1+n_B(\omega_\gamma,T)) G^P(t,t')\right)M^\gamma \nonumber \\ &+e^{i\omega_\gamma (t-t')}M^\gamma \left(n_B(\omega_\gamma,T) G^P(t,t') +(1+n_B(\omega_\gamma,T)) G^N(t,t')\right)M^\gamma \Big]
\end{align}

Now putting this in the equation of motion of the fermion operator of system of interest in \eqref{second_eom},
\begin{align} \label{third_eom}
 |\dot{c}\rangle =&-iH|c\rangle-i\sum_\beta   |\eta^{ph}_{\beta}\rangle - i\sum_\beta  \int_{t_0}^{t} dt' \, \Sigma^{r,ph}_{\beta}(t-t') |c(t')\rangle \nonumber \\ & -i  |\eta_L\rangle - i  \int_{t_0}^{t} dt' \, \Sigma^r_{L}(t-t') |c(t')\rangle-i  |\eta_R\rangle - i  \int_{t_0}^{t} dt' \, \Sigma^r_{R}(t-t') |c(t')\rangle
\end{align}
We know from the standard Langevin equation analysis that once we have all the noise and damping terms we can find the time dependent inscattering matrices \cite{Dhar_PhysRevB.73.085119}. 
\begin{align}
\Gamma^{ph}(t,t')&=\sum_\gamma M^\gamma \Big(
e^{-i\omega_\gamma (t-t')}n_B(\omega_\gamma,T) G^N(t,t') 
+e^{-i\omega_\gamma (t-t')}(1+n_B(\omega_\gamma,T)) G^P(t,t') \nonumber \\
&\qquad\qquad\qquad
+e^{i\omega_\gamma (t-t')}n_B(\omega_\gamma,T) G^P(t,t') 
+e^{i\omega_\gamma (t-t')}(1+n_B(\omega_\gamma,T)) G^N(t,t')
\Big)M^\gamma.
\end{align}

The in-scattering (lesser) self-energy due to phonons is defined from the
noise correlator as,
\begin{equation}
\Sigma^{<,ph}_{ij}(t,t')
=i \sum_\gamma\langle \eta^{\dagger,ph}_{\gamma,j}(t') \eta^{ph}_{\gamma,i}(t) \rangle 
\label{eq:Sigma_less_noise_def}
\end{equation}
Substituting \eqref{phonon_noise} and expanding, we obtain
\begin{align}
\Sigma^{<,ph}_{ij}(t,t')
&\approx i\sum_{\gamma}\sum_{mn}
   M^\gamma_{im} M^\gamma_{nj}
   \Big[
     \big\langle C^{\,ph\,\dagger}_{n\gamma}(t')
     C^{\,ph}_{m\gamma}(t)\big\rangle
    +\big\langle K^{\,ph\,\dagger}_{n\gamma}(t') C^{\,ph}_{m\gamma}(t)\,\big\rangle
    \nonumber\\
&\hspace{2.8cm}
    +\big\langle C^{\,ph\,\dagger}_{n\gamma}(t')K^{\,ph}_{m\gamma}(t)\,\big\rangle
    +\big\langle K^{\,ph\,\dagger}_{n\gamma}(t')
    K^{\,ph}_{m\gamma}(t)\, \big\rangle
   \Big].
\label{eq:Sigma_less_eph_expanded}
\end{align}
Each average factorizes into an electronic and a phononic part. For example,
\begin{align}
\big\langle
C^{\,ph\,\dagger}_{n\gamma}(t') C^{\,ph}_{m\gamma}(t)\,
           \big\rangle
&= \big\langle c_n^\dagger(t')b_\gamma^\dagger(t')\,b_\gamma(t)\; \,c_m(t)\big\rangle
\simeq \big\langle c_n^\dagger(t') c_m(t)\big\rangle\,
       \big\langle b_\gamma^\dagger(t') b_\gamma(t)\big\rangle,
\end{align}
and similarly for the other three terms. Using the definitions,
\begin{align}
G^N_{m,n}(t,t') &= \langle c_n^\dagger(t') c_m(t)\rangle, &
G^P_{m,n}(t,t') &= \langle c_m(t) c_n^\dagger(t')\rangle,\\
B^{ab}_{\gamma\gamma}(t,t') &= \langle b_\gamma^\dagger(t') b_\gamma(t)\rangle, &
B^{em}_{\gamma\gamma}(t,t') &= \langle b_\gamma(t) b_\gamma^\dagger(t')\rangle,
\end{align}
one finds that the noise correlator can be written compactly as
\begin{equation}
\Sigma^{<}_{ph}(t,t')
= i \sum_\gamma
    M^\gamma \Big[ G^N(t,t')\, B^{em}_{\gamma\gamma}(t,t')
                  +G^N(t,t')\, B^{ab}_{\gamma\gamma}(t',t) \Big] M^\gamma.
\label{eq:Sigma_less_eph_BGN}
\end{equation}
Taking the equilibrium form for the Boson operators like earlier,
 the explicit form of
the e-ph in-scattering self-energy,
\begin{align}
\Sigma^{<}_{ph}(t,t')
&= i \sum_\gamma
    M^\gamma \Big[
      \big(1+n_B(\omega_\gamma,T)\big)\, G^N(t,t')\, e^{-i\omega_\gamma (t-t')}
      +n_B(\omega_\gamma,T)\, G^N(t,t')\, e^{+i\omega_\gamma (t-t')}
    \Big] M^\gamma.
\label{eq:Sigma_less_eph_final_time}
\end{align}
Finally to obtain the steady state results we use the Fourier transform similar to the study \cite{Dhar_PhysRevB.73.085119} and use the standard lesser/greater Green's functions
$G^{</>}$, $G^<(t,t')=iG^N(t,t')$ and $G^>(t,t')=-iG^P(t,t')$, the same
expression can be cast in the more familiar form given in the section \ref{sec:NEGF_Equations}.
\begin{align} \label{broadening_appendix}
    \Gamma_{\text{ph}}(E)=i\sum_\gamma &M^\gamma \big{(}-n_B(\omega_\gamma,T) G^<(E-\omega_\gamma) +(1+n_B(\omega_\gamma,T)) G^>(E-\omega_\gamma)
    \nonumber
    \\
    &+n_B(\omega_\gamma,T) G^>(E+\omega_\gamma) -(1+n_B(\omega_\gamma,T)) G^<(E+\omega_\gamma)\big{)}M^\gamma
\end{align}
\begin{align}\label{inscattering_appenidx}\Sigma_{\text{ph}}^<(E)=\sum_\gamma M^\gamma \left(n_B(\omega_\gamma,T) G^<(E-\omega_\gamma) +(1+n_B(\omega_\gamma,T)) G^<(E+\omega_\gamma)\right)M^\gamma
	\end{align}

	
		\section{Approximation of equilibrium e-ph bath\cite{Fransson_PhysRevB.102.235416}}  \label{fransson_approx}
        
\begin{figure}[b]
		\centering 
        \subfigure []{\includegraphics[width=0.32\linewidth,height=0.27\linewidth]{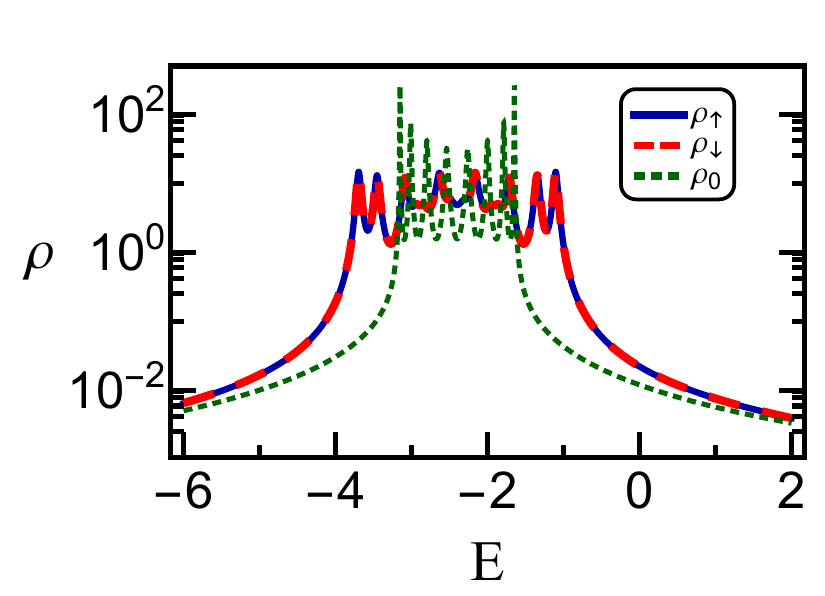}}
\subfigure []{\includegraphics[width=0.32\linewidth,height=0.27\linewidth]{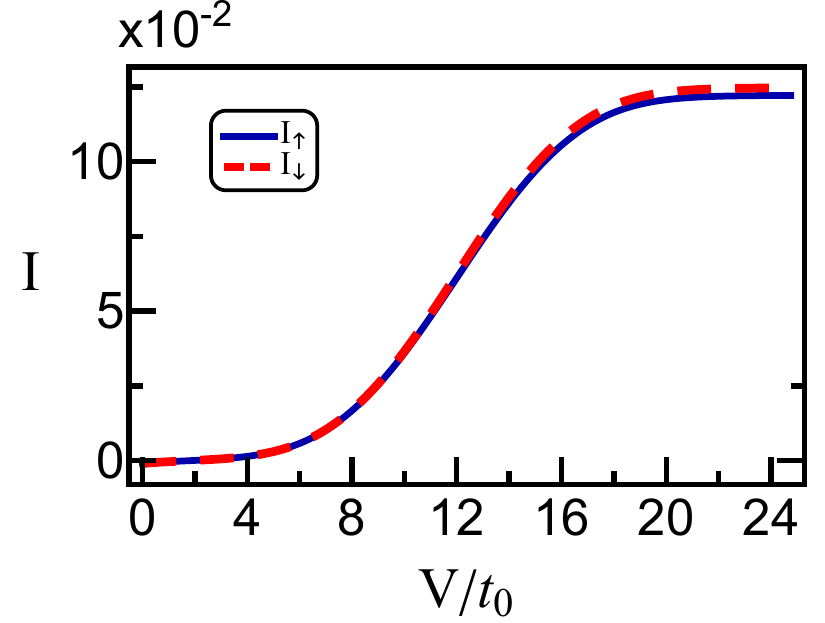}}
        \subfigure []{\includegraphics[width=0.32\linewidth,height=0.27\linewidth]{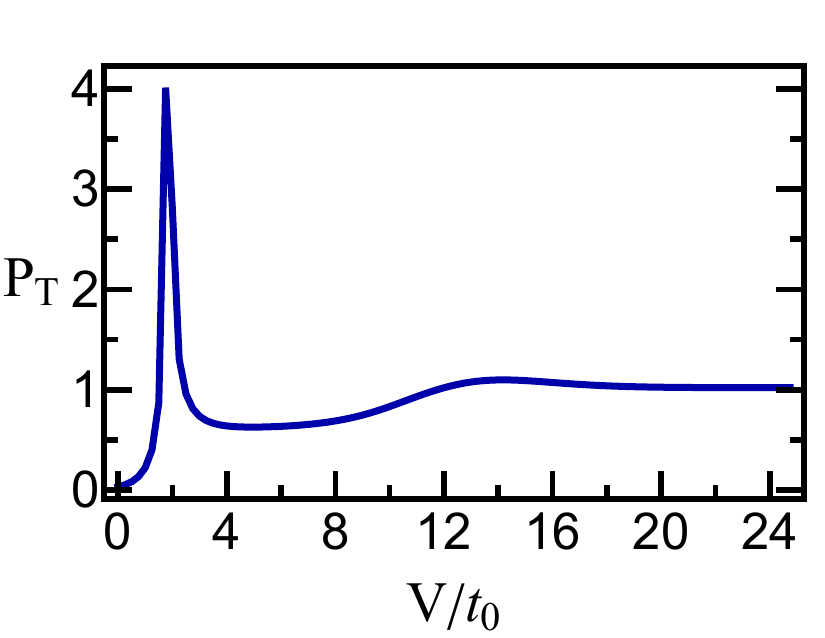}}
        \caption{
                \textbf{(a)} Energy-dependent density of states obtained using the approximations introduced in Appendix \ref{fransson_approx}, \textbf{(b)} Current-voltage characteristics for opposite left-lead magnetizations. \textbf{(c)} Spin polarization as a function of bias voltage. The parametes are the same as given in Fig. \ref{ideal_fig}}
		\label{Fransson_method_fig}
	\end{figure}
            This involves simplifying the broadening matrix. To do this it is assumed that the Green's Functions at energies $E\pm\omega$ can be approximated by their equilibrium values. Further these equilibrium values are just taken by considering the on site energy $\varepsilon_0$. This means that the energy shifted Green's Functions are just considered as, $G^<(E)=  2i \pi\delta(E-\epsilon_0)f_0(\epsilon_0), G^>(E)=  -2i \pi\delta(E-\epsilon_0)(1-f_0(\epsilon_0))$. Using this in the expression for broadening matrix,
            \begin{align}
    \Gamma_{\text{ph}}(E)=2\pi &M_0^2 \big{[}n_B(\omega_0) f_0(\varepsilon_0) \delta(E-\omega_0-\varepsilon_0) +(1+n_B(\omega_0))(1- f_0(\varepsilon_0)) \delta(E-\omega_0-\varepsilon_0)
    \nonumber
    \\
    &+n_B(\omega_0) (1-f_0(\varepsilon_0)) \delta(E+\omega_0-\varepsilon_0) +(1+n_B(\omega_0))f_0(\varepsilon_0) \delta(E+\omega_0-\varepsilon_0)\big{]}
\end{align}
We now replace the Dirac delta peaks by Lorentzians of finite width $\eta$,
\begin{equation}
\delta(x)\ \longrightarrow\ \delta_\eta(x)\equiv \frac{1}{\pi}\frac{\eta}{x^{2}+\eta^{2}},
\qquad \eta>0,
\end{equation}
so that $\lim_{\eta\to 0^+}\delta_\eta(x)=\delta(x)$. Applying this to the phonon broadening yields,
\begin{align}
\Gamma_{\text{ph}}(E)\ \approx\ 2\pi M_0^2 \Big[
&\big(1-f_0(\varepsilon_0)+n_B(\omega_0)\big)\,\delta_\eta(E-\varepsilon_0-\omega_0)
\nonumber\\
&+\big(f_0(\varepsilon_0)+n_B(\omega_0)\big)\,\delta_\eta(E-\varepsilon_0+\omega_0)
\Big].
\end{align}
 Now the PV integral of the Lorentzian is,
 \begin{equation}
\,\mathcal{P}\!\int_{-\infty}^{\infty} dE'\;
\left(\frac{1}{\pi}\frac{\eta}{E'^2+\eta^2}\right)\frac{1}{E-E'}
=\frac{E}{E^2+\eta^2}.
\end{equation}

Using the above identity in the following definition,
\begin{align}
\Sigma^R(E)=
 \frac{1}{2\pi}\mathcal{P}\!\int_{-\infty}^{\infty}\! dE'\,
\frac{\Gamma_{\mathrm{ph}}(E')}{E-E'}
-\frac{i}{2}\Gamma_{\mathrm{ph}}(E)=M_0^2\Sigma(E) 
\end{align}
where,
\begin{equation}
		\Sigma(E) = \frac{n_B(\omega_0) + 1 - f_0(\varepsilon_0)}{E - \omega_0 - \varepsilon_0 + i\eta} 
		+ \frac{n_B(\omega_0) + f_0(\varepsilon_0)}{E + \omega_0 - \varepsilon_0 + i\eta}.
	\end{equation}
	where $\eta$ is a parameter used for denoting internal broadening due to electron phonon interaction, $\eta=\frac{t_0}{4}$. This is the simplification assumed in the analysis by Fransson in various studies \cite{Fransson_PhysRevB.102.235416,Fransson_What_it_CISS_https://doi.org/10.1002/ijch.202200046}. Considering the above approximations, we get the following expression for the retarded Green's Function at energy E, \cite{Datta_2005,Fransson_PhysRevB.102.235416},
    \begin{equation}
		G^R(E) = \left(E - H - \Sigma(E) M_0^2 + \frac{i}{2} (\Gamma^{L} + \Gamma^{R}) \right)^{-1},
	\end{equation}
 Additionally the e-ph contribution to the density of occupied states is given as,
\begin{align}
\Sigma_{\mathrm{ph}}^{<}(E)
=&
2\pi i\,M_0^{2}\,f_0(\varepsilon_0)
\Big[
n_B(\omega_0)\,\delta(E-\varepsilon_0-\omega_0)
+\big(1+n_B(\omega_0)\big)\,\delta(E-\varepsilon_0+\omega_0)
\Big]\nonumber \\&\approx
2\pi i\,M_0^{2}\,f_0(\varepsilon_0)
\Big[
n_B(\omega_0)\,\delta_\eta(E-\varepsilon_0-\omega_0)
+\big(1+n_B(\omega_0)\big)\,\delta_\eta(E-\varepsilon_0+\omega_0)
\Big].
\end{align}
    So the density of occupied electronic states is given as,
    \begin{align}
G^<(E)=i\left(f_LG^R\Gamma_LG^A+f_RG^R\Gamma_R G^A \right)+G^R(E)\Sigma_{\mathrm{ph}}^<(E)G^A(E)
\end{align}
By similar calculation,
\begin{align}
\Sigma_{\mathrm{ph}}^{>}(E)\approx
-2\pi i\,M_0^{2}\,(1-f_0(\varepsilon_0))
\Big[\big(1+n_B(\omega_0)\big)\,\delta_\eta(E-\varepsilon_0-\omega_0)
+n_B(\omega_0)\,\delta_\eta(E-\varepsilon_0+\omega_0)
\Big].
\end{align} and,
    \begin{align}
G^>(E)=-i\left((1-f_L)G^R\Gamma_LG^A+(1-f_R)G^R\Gamma_R G^A \right)+G^R(E)\Sigma_{\mathrm{ph}}^>(E)G^A(E)
\end{align}
    \subsection{Current and density analysis} \label{Fransson_results}
We next present the results obtained by analysing the model using the approximations discussed above, shown in Fig. \ref{Fransson_method_fig}. In Fig. \ref{Fransson_method_fig}(a), we observe that, similar to the fully self-consistent analysis, the density of states spreads out in the presence of electron-phonon interactions. The inclusion of interactions leads to a redistribution of spectral weight over a wider energy range, indicating that inelastic processes again play an important role in shaping the electronic structure.
However, there is a significant qualitative difference compared to the self-consistent case. In particular, the individual resonant peaks in the DOS remain clearly visible when the approximations are employed, in contrast to the strongly smeared spectrum obtained in the self-consistent treatment. This difference highlights the sensitivity of the spectral properties to the level of approximation used in treating electron-phonon interactions.

The corresponding particle current is shown in Fig. \ref{Fransson_method_fig}(b). As compared to the self-consistent case, the overall magnitude of the current remains small. However, for larger voltages we can see a visible distinction between the two currents at opposite lead magnetization.
Finally, we plot the spin polarization as a function of bias voltage in Fig. \ref{Fransson_method_fig}(c). Unlike the self-consistent analysis, a finite polarization is observed within this approximation scheme. This indicates that the use of the above approximations leads to qualitatively different conclusions regarding the emergence of spin-selective transport, despite the overall current remaining small.
\end{widetext}
   \bibliography{Refer}
\end{document}